\documentclass[annual]{acmsiggraph}

\TOGonlineid{}

\TOGvolume{}
\TOGnumber{}

\TOGarticleDOI{}

\TOGprojectURL{}
\TOGvideoURL{}
\TOGdataURL{}
\TOGcodeURL{}

\title{A Reciprocal Formulation of Nonexponential Radiative Transfer.   2: Monte Carlo Estimation and Diffusion Approximation}

\author{ Eugene d'Eon \\
8i }

\pdfauthor{ }


\usepackage{smrdefaults}

\usepackage{color}

\usepackage{multirow}

\usepackage{rotating}
\newcommand{\new}[1]{#1}
\newcommand{\remove}[1]{}

\newcommand{\newtwo}[1]{#1}
\newcommand{\removetwo}[1]{}

\newcommand{\newthree}[1]{#1}

\newcommand{\mfp}{\langle s \rangle}

\newcommand{\s}{\langle s_c \rangle}  
\renewcommand{\ss}{\langle s_c^2 \rangle}  
\newcommand{\sss}{\langle s_c^3 \rangle}  
\newcommand{\ssss}{\langle s_c^4 \rangle}  

\newcommand{\su}{\langle s_u \rangle}  
\newcommand{\ssu}{\langle s_u^2 \rangle}

\newcommand{\vect}[1]{\vec{#1}}

\newcommand{\phipt}{\phi_{\text{pt}}}

\newcommand{\F}{\mathcal{F}_d}    \renewcommand{\Finv}{\mathcal{F}_d^{-1}}  \newcommand{\Finvthree}{\mathcal{F}_3^{-1}}  

\renewcommand{\P}{\mathcal{P}}

\newcommand{\e}{\text{e}}

\newcommand{\pos}{\mathbf{x}}

\newcommand{\dir}{\vect{\omega}}

\def\rm{R^{-}}

\usepackage{paralist}
\usepackage{subfigure}
\usepackage{rotating}
\usepackage{amsgen}
 
\usepackage{alltt}
\usepackage{ulem}
\usepackage{xcolor}

\begin{document}

\normalem

\maketitle

\begin{abstract}
  When lifting the assumption of spatially-independent scattering centers in classical linear transport theory, collision rate is no longer proportional to angular flux / radiance because the macroscopic cross-section\remove{s} $\Sigma_t(s)$ depends on the distance $s$ to the previous collision or boundary.  \new{This creates a nonlocal relationship between collision rate and flux and requires revising a number of familiar deterministic and Monte Carlo methods.}  We generalize collision and track-length estimators\remove{, and other related Monte Carlo methods,} to support unbiased estimation of either flux integrals or collision rates in generalized radiative transfer (GRT).  \remove{We consider collision estimators with additional fictitious scattering that, in the limit of infinite fictitious density, become track-length estimators.  The generalizations support distinct correlated- and uncorrelated-origin free-path statistics, which are required to imbed non-exponential transport in bounded scenes in a reciprocal manner.}
To provide benchmark solutions for the Monte Carlo estimators, we derive the four Green's functions for the isotropic point source in infinite media with isotropic scattering.  Additionally, new moment-preserving diffusion approximations for these Green's functions are derived, \newtwo{which} \new{ reduce to algebraic expressions involving the first four moments of the  free-path lengths between collisions}\remove{, and their accuracy is evaluated for power-law free-path distributions.  The general imbalance of flux and collision rate is explored in a variety of scenarios and we highlight new subtleties regarding the estimation of reaction rates in classical anisotropic random media and GRT}.
\end{abstract}

\keywordlist

\section{Introduction}

  In the study of particle transport~\cite{williams71,davison57}, light transport~\cite{chandrasekhar60,arridge99,marshak05,pharr16}, seismic wave propagation~\cite{sato2012seismic}, \new{heat transfer~\cite{modest03}}, and other fields~\cite{jagers75,duderstadt79}, exact solutions of the transport equation are not possible in most cases and estimations based on Monte Carlo and diffusion theory are common.  The nascent theory of \emph{generalized radiative transfer} (GRT) extends the transport equation to support spatial correlation between collisions, and this requires Monte Carlo and deterministic methods to be likewise generalized.

  This paper presents new methods for estimating collision rates and flux integrals in monoenergetic, time-independent GRT.  These methods are organized into three mains groups:
  \begin{compactitem}
    \item (Section~\ref{sec:MC}) Collision and track-length estimators for piecewise-homogeneous systems in $\mathbb{R}^d$ with general anisotropic scattering
    \item (Section~\ref{sec:greens}) Green's functions for the isotropic point source in infinite homogeneous media in $\mathbb{R}^d$ with isotropic scattering
    \item (Section~\ref{sec:diffusion}) Two families of moment-preserving diffusion approximations for the Green's functions in Section~\ref{sec:greens} plus an additional pair of diffusion approximations for collision rate with a general anisotropic scattering law.
  \end{compactitem}
  The Monte Carlo estimators and Green's functions are presented together in order to cross-validate each other, and the diffusion approximations follow directly from the Green's functions.  All of the methods reduce to familiar forms in the classical case of no correlation and are all motivated by two unique properties of nonclassical transport: nonlocal interaction probabilities and the distinct statistics for free path lengths when leaving a collision versus a deterministic location.

  \paragraph{Nonlocal Interaction}

    Classical linear transport includes a principle of local interaction~\cite{preisendorfer65,grant1969discrete} that follows from the assumption of independent scattering centers.  This guarantees that vector collision rate density and radiance (angular flux) are always proportional.  The same proportionality holds for the scalar collision rate density and fluence.  Classical Monte Carlo estimators and deterministic approximations for flux and collision rate are therefore equivalent up to a constant factor (the macroscopic cross section $\Sigma_t$). 

    In GRT, a two-point correlation between collisions includes a memory of the distance $s$ to the previous collision, which appears in the cross section $\Sigma_{tc}(s)$~\cite{larsen2007,larsen11}.  This non-Markovian property breaks the local proportionality of collision rate density and flux~\cite{deon14,mulatier2014universal}.  We find that moment-preserving diffusion approximations, Green's functions and Monte Carlo estimators therefore require distinct forms in GRT, depending upon whether or not the desired physical quantity to estimate is proportional to the collision rate or to the flux.

  \paragraph{Dual Statistics}

\begin{figure}
      \centering
      \includegraphics[width=1.06 \linewidth]{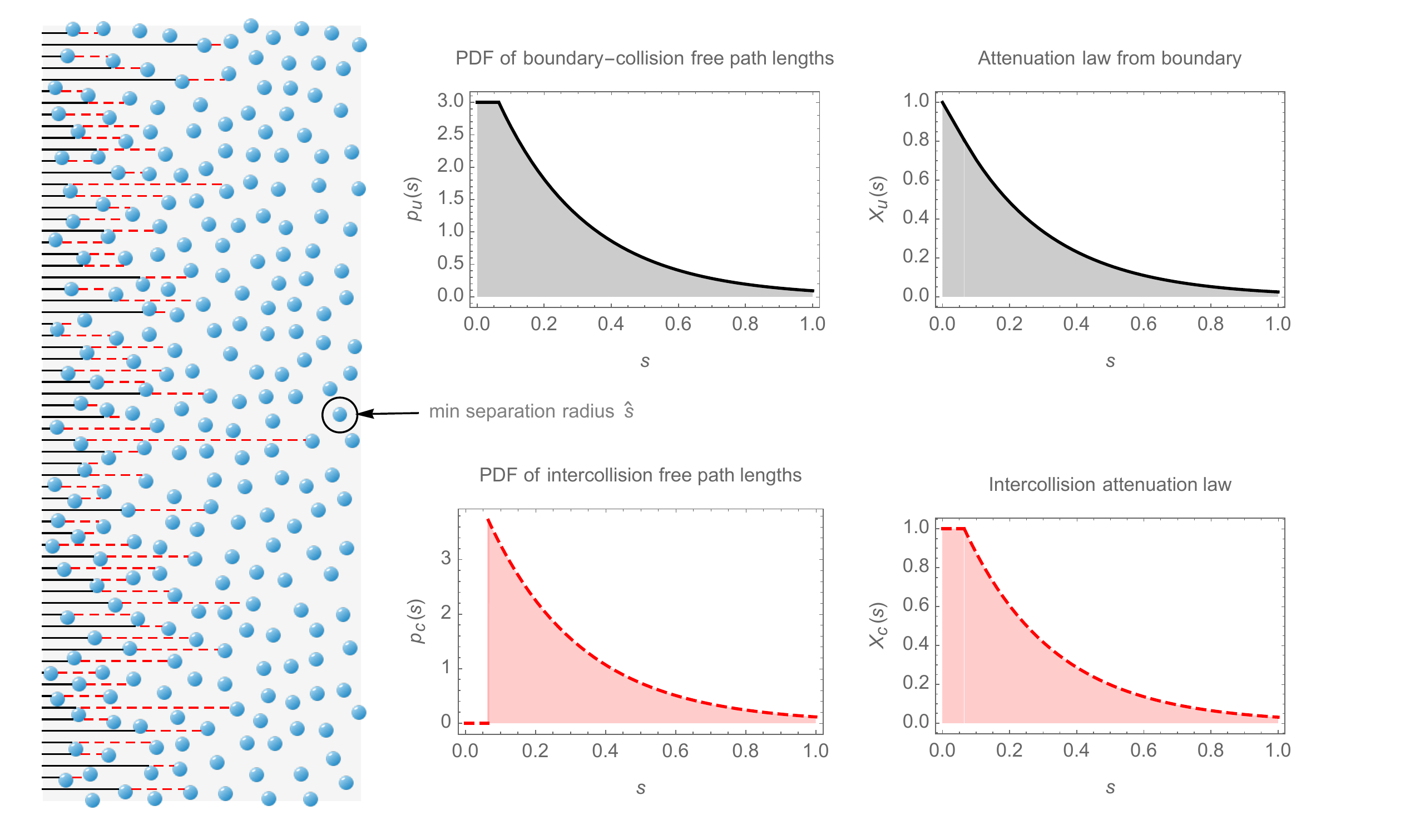}
      \caption{When scatterers in a random medium are spatially correlated, the free-path length statistics between collisions are necessarily distinct from those for paths beginning at a boundary interface.  Here we illustrate the case of negatively-correlated convex scatterers separated by a minimum distance $\hat{s} = 0.065$.  For paths beginning at the left boundary of a unit thickness slab (solid-black) collisions can occur arbitrarily close to the boundary and the related path length PDF $p_u(s)$ and attenuation law $X_u(s)$ reflect this.  Continuing in the same direction from the first collision to the second collision (red-dashed), we find path lengths with a minimum length $\hat{s}$.  The intercollision free-path distribution $p_c(s)$ is therefore identically zero for $s < \hat{s}$ due to the scatterers separation, and the attenuation law between collisions $X_c(s)$ is $1$ for this initial distance.}
      \label{fig-pcpupretty} 
    \end{figure}

    Multiple scattering in a bounded homogeneous system with spatial correlation exhibits two distinct classes of free-path statistics: those for the free-path lengths between collisions, and those for paths originating at a deterministic location, such as on a boundary~\cite{audic1993monte}.  From these statistics, a pair of distinct attenuation laws for the system immediately follow~\cite{deon2018reciprocal}.  Intuition for the distinction between these statistics is easily found by considering a simple slab of well-separated particles (Figure~\ref{fig-pcpupretty}).  We use the label `c' for correlated origins and `u' for uncorrelated (deterministic) origins and denote the two free-path distributions $p_c(s)$ and $p_u(s)$, with attenuation laws $X_c(s)$ and $X_u(s)$ (for further details, which we summarize in Table~\ref{tab:notation}, see \cite{deon2018reciprocal}).

    To account for dual statistics in GRT, both $\Sigma_{tc}(s)$ and $\Sigma_{tu}(s)$ can appear in the new Monte Carlo estimators, depending on the statistical correlation of the previous path vertex to the scatterers.  Similarily, two classes of emission lead to distinct source terms: $Q_c(\pos)$ for emission from the correlated microstructure, and $Q_u(\pos)$ for uncorrelated emission, which produce distinct Green's functions and diffusion approximations depending on the spatial correlation of the point source to the scatterers. In constrast to classical transport, which exhibits effectively only one point source Green's function (up to a constant), in GRT we find four (correlated/uncorrelated, collision-rate/fluence).
    \begin{table*}[t]
    \centering   

    \scalebox{0.7}{
    \begin{tabular}{| c | l | c |} 
    
      \hline    
      \vline width0pt height2.3ex
      \textbf{Symbol} & \textbf{Description} & \textbf{Relations} \\
\hline \multicolumn{3}{|c|}{ \small \emph{medium-\textbf{c}orrelated \new{(stochastic)} free path origins }} \\
      \hline
      $s$ & {\small distance since last medium collision or correlated birth} & \\
      \hline
      $\Sigma_{tc}(s)$ & {\small correlated macroscopic cross section} & $\Sigma_{tc}(s) = \frac{p_c(s)}{X_c(s)}$ \\
      \hline
      $p_c(s)$ & {\small correlated free-path distribution} & $p_c(s) = \Sigma_{tc}(s) \e^{-\int_0^s \Sigma_{tc}(s') ds'} = - \frac{\partial}{ \partial s} X_c(s) = \s \frac{\partial^2}{ \partial s^2} X_u(s)$ \\
      \hline
      $X_c(s)$ & {\small correlated-origin transmittance} & $X_c(s) = 1 - \int_0^s p_c(s') ds'$ \\
      \hline
      $\mfp_c$ & {\small mean correlated free-path} & $\mfp_c = \int_0^\infty p_c(s) \, s \, ds$ \\
      \hline
      $\ss$ & {\small mean squared correlated free-path} & $\ss = \int_0^\infty p_c(s) \, s^2 \, ds$ \\
      \hline
\multicolumn{3}{|c|}{ \small \emph{medium-\textbf{u}ncorrelated \new{(deterministic)} free path origins }} \\
      \hline
      $s$ & {\small distance since last surface/boundary or uncorrelated birth} & \\
      \hline
      $\Sigma_{tu}(s)$ & {\small uncorrelated macroscopic cross section} & $\Sigma_{tu}(s) = \frac{p_u(s)}{X_u(s)}$ \\
      \hline
      $p_u(s)$ & {\small uncorrelated (equilibrium) free-path distribution} & $p_u(s) = \Sigma_{tu}(s) \e^{-\int_0^s \Sigma_{tu}(s') ds'} = - \frac{\partial}{ \partial s} X_u(s) = \frac{X_c(s)}{\mfp_c}$ \\
      \hline
      $X_u(s)$ & {\small uncorrelated-origin transmittance} & $X_u(s) = 1 - \int_0^s p_u(s') ds'$ \\
      \hline
      $\su$ & {\small mean uncorrelated free-path} & $\su = \int_0^\infty p_u(s) \, s \, ds$ \\
      \hline

    \end{tabular} 
    }   
    \caption{\label{tab:notation}Summary of our notation and relationships between quantities in GRT.}
  \end{table*}

  \paragraph{Motivation and Approach}

    The application of infinite medium Green's functions to solve linear transport problems in bounded domains has seen broad use, especially in tissue optics~\cite{farrell92,kienle97} and computer graphics~\cite{jensen01,donner05}.  Deterministic solutions to model problems also find use in biased acceleration methods~\cite{fleck84} or unbiased importance-sampling and guiding of Monte Carlo methods towards zero-variance estimation~\cite{dwivedi82,hoogenboom08b}.  The deterministic methods in this paper are derived with the future goal of extending these approaches to materials where the transport is better represented by the nonexponential random flights of GRT.

    From the generalized Peierls's integral equation for nonexponential random flights~\cite{grosjean51,larsen11} we derive exact point-source Green's functions using a Fourier transform approach~\cite{grosjean51,zoia11e,deon14}.  For the case of Gamma-2 flights\footnote{Several works~\cite{frank15,vasques2016nonclassical,makine2018exact} have noted that Gamma-2 random flights produce diffusion solutions in 3D.  However, by applying a local/classical conversion from collision rate to fluence, which does not hold in GRT, their conclusions about the fluence are approximate.} in 3D, we find analytic Green's functions in the point- and plane-source cases, and these produce the first exact GRT benchmark solutions (which play an important role in linear transport theory~\cite{ganapol08}).  For most other correlations, however, the Fourier inversion cannot be performed analytically, and the efficiency of the exact solutions is then limited.  For these cases, we also derive compact diffusion approximations.  

    In forming diffusion approximations, there is a degree of arbitrariness based on the desired physical quantities to be preserved~\cite{larsen10}.  We choose a moment-preserving approach such that the first two even spatial moments of selected scalar quantities are exactly preserved by the approximations.  We derive multiple approximations for both the fluence $\phi(r)$ and the collision rate density $C(r)$, considering both uncorrelated and correlated emission.  In the case of collision rate density, the mean number of collisions $\langle r^0 C(r) \rangle$ and mean-square distance of collision $\langle r^2 C(r) \rangle$ are exactly preserved.  For the fluence derivations, the total energy in flight and mean square distance (from the source) of energy in flight are simultaneously preserved by exactly preserving $\langle r^0 \phi(r) \rangle$ and  $\langle r^2 \phi(r) \rangle$.  

    For each of the four Green's functions, we derive two diffusion approximations: the classical $P_1$ diffusion approximation, where all orders of scattering are accounted for by a single diffusion mode, and also a hybrid modified-diffusion approximation due to Grosjean~\shortcite{grosjean56a} that separates out the exact uncollided term and represents the remainder with a diffusion mode.  This specific form of modified diffusion was chosen for its increased performance for high absorption levels and near sources, and has proven useful in the classical case when rendering tissue in computer graphics~\cite{deon11a,deon14dual} and predicting the appearance of 3D-printed materials~\cite{papas2013fabricating}.

    For piecewise-homogeneous media with general scattering laws, we also generalize the collision and track-length estimators for GRT.  Our approach follows Spanier et al.~\shortcite{spanier69,spanier66}, beginning with the definition of the analog random walk for GRT in bounded scenes.  We then generalize the collision estimators to permit a general class of fictitious scattering that, in the limit of infinite fictitious density, produces the track-length estimators for either collision rates or flux integrals.  We find that separate tallies are required for collision rates and fluxes, and that the collision and track-length estimators can be used together during the same random walk to estimate both quantities simultaneously.  Nonanalog walks, implicit capture and next-event estimation are also discussed.

  \subsection{Related Work}

    In this paper\footnote{Our use of ``GRT'' follows Davis et al.~\shortcite{davis14}, but we apply it to a broader class of reciprocal transport processes with general intercollision statistics.} we use ``GRT'' to refer specifically to any linear non-Beerian transport process that discards the notion of exponential free-path distributions (FPDs) characterized only by a mean, and instead forms semi-Markov random flights using the ensemble-averaged two-point intercollision free-path statistics exhibited by a given random medium~\cite{grosjean51,alt1980biased,audic1993monte,peltoniemi1993radiative,davis06,larsen2007,moon07,taine2010generalized,larsen11,davis14}.  We intentionally exclude higher-order memory~\cite{myneni91}, stochastic absorption and phase functions~\cite{jarabo18} and heterogeneous mean densities~\cite{bitterli2018radiative}, all of which would significantly complicate some of the present derivations.

    \paragraph{Motivation for GRT}

      The primary goal of GRT is to extend linear transport methodology to cover a wider class of participating media while avoiding the significant challenge of exactly solving a fully stochastic equation of radiative transfer~\cite{frisch1968wave,papanicolau1971stochastic,ishimaru78,borovoi2006multiple}.  In stochastic radiative transfer, the medium properties become random variables in order to statistically account for unresolved spatial and temporal fluctuations that inevitably arise in any realistic system~\cite{williams1974random}.  The desired transport quantities to estimate are then the mean collision rates and fluxes over all possible realizations of the medium.  Homogenization to an effective classical medium with exponential attenuation is only accurate when the magnitude of such fluctuations is small and the correlation lengths are on the order of the wavelength~\cite{ryzhik96,barabanenkov1968radiation,wen1990dense,tsang00}.  In many systems, such as dense suspensions of particles~\cite{ishimaru82,moon07}, clumpy molecular clouds~\cite{boisse1990radiative,witt1996multiple,witt2000multiple}, terrestrial atmospheres~\cite{davis1996scale,marshak1997scale,pfeilsticker1999first,kostinski01}, manufactured materials with random voids~\cite{barthelemy08,svensson13,svensson14}, hot atomic vapors~\cite{mercadier2009levy}, shielding materials~\cite{becker2014measurement} and pebble-bed reactors~\cite{larsen11}, significant deviation from Beer-Lambert attenuation is observed.  The impact of such strong correlations on the mean transport can be severe~\cite{torquato2016hyperuniformity}.

      GRT is an efficient and approximate form of scalar stochastic radiative transfer formed by
      \begin{compactitem}
        \item assuming nonstochastic (independent of $s$) single-scattering albedo $c$ and phase function $P$ 
        \item noting that the mean intercollision free-path statistics $p_c(s)$ averaged over all realizations of the medium is nonexponential~\cite{audic1993monte,moon07,larsen11}
        \item adopting a nonexponential random flight that exactly exhibits $p_c(s)$, and assuming all higher-order correlation negligible.
      \end{compactitem}
      GRT is agnostic to how $p_c(s)$ is determined.  It may be estimated using Monte Carlo sampling within specific realizations of quenched disorder~\cite{audic1993monte,moon07,larsen11}, by deterministic analysis of a periodic structure~\cite{golse12}, or from observed statistics of the motion of organic material~\cite{alt1980biased,kareiva1983analyzing}.  Alternatively, $p_c(s)$ follows from specifying any one of $\Sigma_{tc}(s), \Sigma_{tu}(s), p_u(s), X_c(s), X_u(s)$, by relationships summarized in Table~\ref{tab:notation}.  The mean attenuation law from a deterministic origin $X_u(s)$ may be measured in the laboratory, or calculated from a given stationary random function for $\Sigma_t$~\cite{levermore88,williams1997radiation,kostinski01,davis2011radiation,larsen2014link,frankel2017optical,banko2019stochastic,guo2019fractional}.

    \paragraph{Reciprocity}

      The need for dual statistics in bounded domains was first noted by Audic and Frisch~\shortcite{audic1993monte}, who estimated $p_c(s)$ using Monte Carlo. In the first part of this work~\cite{deon2018reciprocal}, we showed that Helmholtz reciprocity for a diffuse law of reflection in a GRT half space requires the relationship
      \begin{equation}\label{eq:pcXu}
        p_c(s) = \s \frac{\partial^2}{\partial s^2} X_u(s)
      \end{equation}
      where $\s$ is a constant---the mean-free-path length between collisions\footnote{For L{\'e}vy walks with unbounded mean free paths~\cite{liemert2017radiative,binzoni2018generalized}, truncation to a finite mean~\cite{mantegna1994stochastic} is required.}.  This greatly simplifies the specification of media in GRT, bypassing the need for Monte Carlo free-path histograms in \cite{audic1993monte}.  Bitterli et al.~\shortcite{bitterli2018radiative} subsequently confirmed Equation~\ref{eq:pcXu} in an alternative derivation based on mean attenuation laws in continuous random media, noting that $X_u(s)$ is the mean attenuation over all realizations, and $X_c(s)$ is the mean attenuation over only those realizations with a scatterer at the start ($s = 0$).  

      Dual free-path length statistics have also appeared in several other studies.  It was observed in the periodic Lorentz gas~\cite{golse12} and is also the reason why the Milne problem in a binary mixture has two definitions/solutions~\cite{pomraning1989milne}.  The relationship in Equation~\ref{eq:pcXu} corresponds to known relationships between chord lengths and lineal path functions in heterogeneous materials~\cite{torquato93}, and was first suggested in a bounded random flight~\cite{mazzolo09} in order to preserve a remarkable invariance property of chord lengths in convex uniformly-illuminated random media, first proven for exponential media~\cite{bardsley81,blanco2003invariance} and then extended to nonexponential free paths~\cite{mazzolo09,mazzolo2014cauchy}, and more recently observed in the laboratory~\cite{savo2017observation}.

      Without accounting for this distinction in GRT, several related works speak only of ``the attenuation law'' and produce nonreciprocal transport in bounded domains~\cite{davis06,taine2010generalized,davis14,wrenninge17}.  However, the methods in this paper also apply for these formalisms.

    \paragraph{Monte Carlo Methods for Nonclassical Transport}

      Fictitious scattering in nonclassical transport has been previously proposed as an acceleration scheme of the chord-length sampling algorithm in the context of multiphase multidimensional mixtures~\cite{ambos2011statistical}.  This differs from our use of fictitious scattering to transition from collision to track-length estimation in the context of completely general free-path statistics.  

      The classical track-length estimator has also been previously applied for estimating fluence in non-classical transport~\cite{boisse1990radiative,liemert2017radiative}, but not generalized to estimate collision rates.

      Bitterli et al.~\shortcite{bitterli2018radiative} generalized Veach's path-integral formulation of light transport for GRT, manipulating terms to produce a fully reciprocal model suitable to bidirectional estimators.  This formulation is complementary to our present investigations of reaction rate and fluence estimates using track-length estimators.

    \paragraph{Relationship to Stochastic Processes Literature}

      The Monte Carlo and deterministic methods in this paper apply generally to any stochastic processes where a particle moves along straight paths at constant speed between collisions that change the particle's direction and possibly absorb the particle.  This include some, but not all, of the stochastic processes that are studied under names such as L{\'e}vy walks~\cite{uchaikin1997levy,zaburdaev2015levy}, random evolutions~\cite{hersh1974random}, velocity jump process~\cite{othmer2000diffusion} and semi-Markov random flights~\cite{grosjean51,dutka85,majumdar2008universal,zoia11e,caer11,degregorio2012random,degregorio2012flying,pogorui2013random,degregorio2017note}.

    \paragraph{The Generalized Linear Boltzmann Equation}

      In classical linear transport, various equivalent mathematical approaches are loosely divided into two camps~\cite{guth60,wigner1961mathematical} based on whether or not the focus is on the history of a single particle (random flight) or on a single volume element (Boltzmann).  The nonexponential random flights of GRT have been studied since the early 1900s~\cite{grosjean51,dutka85}.  The related \newthree{generalized linear Boltzmann equation (GLBE)} for GRT was more recently proposed and the connection to the integral equation of random flight established~\cite{larsen2007,larsen11}.  Existence and uniqueness of solutions for the GLBE have been established~\cite{frank10} and the GLBE has been extended to support anisotropic random correlated media~\cite{vasques13} with direction-dependent cross sections.  Other Boltzmann-like equations have appeared~\cite{peltoniemi1993radiative,golse12,rukolaine2016generalized,binzoni2018generalized}, and some equivalences between them have been established~\cite{larsen2017equivalence}.  The most relevant aspect of these works to the present is not the integro-differential form of the transport equation, however, but the formalism of total macroscopic cross sections $\Sigma_t(s)$ with a free-path-length memory variable $s$.  Through this lens, nonexponential random flights take on a familiar linear-transport character, and the subtleties of collision rate and flux balance and generalization of familiar Monte Carlo methods become more evident.

    \paragraph{Diffusion Asymptotics for Nonclassical Transport}

      The Green's function derivations and related moment-preserving diffusion approximations in this paper expand upon our previous work~\cite{deon14} by including the case of uncorrelated emission, and by performing a completely general moment analysis in the Fourier domain that avoids the previous need for Fourier inversions.  In addition to proving conjectures from this previous work, we also prove new results about the diffusion coefficient's dependence on spatial dimension $d$. The relationship to other diffusion asymptotics derived from a parameter-of-smallness approach~\cite{frank10,larsen11} is also discussed.

      We do not presently consider diffusion approximations that preserve the rigorous ``asymptotic'' diffusion length $\kappa_{\text{tr}}$, which may be desired in deep-penetration and shielding applications.  For expanded approximations in classical transport that preserve both multiple even spatial moments and $\kappa_{\text{tr}}$ see \cite{grosjean63a,grosjean63b} and \cite{larsen10}.  For a $\kappa_{\text{tr}}$-preserving diffusion approximation of collision rate density under correlated emission in GRT, see~\cite{deon14}.  Also, alternatives to diffusion, such as saddlepoint methods~\cite{gatto2017saddlepoint}, may be preferred in some cases, but are not treated presently.

    \paragraph{Relationship to Wave Scattering Theories}

      Theories for wave propagation in random media are closely related to the scalar equations of radiative transfer (RT) that neglect polarization and interference aspects~\cite{keller1962wave,ishimaru1991wave,apresyan1996radiation}.  Originally criticized for being phenomenological, more recently RT has enjoyed a strengthened relationship to ``honest'' wave formalisms~\cite{mishchenko08}, a relationship that continues to evolve and surprise~\cite{pierrat2014invariance}.  Wave theories for stochastic media have considered the influence of correlation and have proposed analogous concepts to those in scalar GRT.  The recognition of the non-Markovian property of wave transport in spatially-correlated random media and the suggestion of two-point memory as a practical simplification goes back at least as far as Hoffman~(\shortcite{hoffman1964wave}, p.137).  Also, in the discrete picture of a random medium, the observation of the nonlocal relationship between total and exciting fields is similar to the nonlocal proportionality of collision rate density and fluence in GRT; ``what relation exists between the exciting field acting on a scatterer at a point, and the total field which would exist at that point if the scatterer were not there"~\cite{waterman1961multiple}.  The early theories of Foldy and Lax assumed a constant proportionality, but correlation exists in any realistic medium and the exact relationship between the exciting and total fields is that of an operator (\cite{frisch1968wave} p.172).

  \subsection{Outline}

  The paper is structured into three main theoretical sections and concludes with a discussion.  In Section~\ref{sec:balance} we examine the nonlocal proportionality between collision rate and flux that motivates the following theory.  In Section~\ref{sec:MC}, we generalize the collision and track-length estimators for GRT.  In Section~\ref{sec:greens}, we derive the infinite medium Green's functions for isotropic point sources with isotropic scattering in a monoenergetic steady-state setting.  The Green's functions and Monte Carlo estimators are used to cross validate each other in Section~\ref{sec:validation}.  Related moment-preserving diffusion approximations are derived in Section~\ref{sec:diffusion}.   A number of related (but untested) details relating to Monte Carlo estimation in GRT are included in Appendix~\ref{appendix:MC}.

\section{The Balance of Flux and Collision Rate}\label{sec:balance}

  \newtwo{To motivate the following sections, we begin by recalling the relationship between}\removetwo{In this section we study the relationship between} collision rate and flux in both classical and generalized time-independent linear transport theories.  \removetwo{New complexities arise in anisotropic random media and GRT, requiring generalization of standard practice.}  We denote the \emph{collision rate density}\footnote{\cite{liemert2017radiative} use ``radiance'' for vector collision rate density, which risks causing confusion with respect to the density of particles in flight.} $C(\pos,\dir)$, where
   \begin{align}\label{eq:Cdefn}
    C(\pos,\dir) dV d\dir = \, &\text{rate at which particles in $dV d\dir$ } \\
                             &\text{about $(\pos,\dir)$ enter collisions } \nonumber
   \end{align}
   and denote its scalar counterpart, the \emph{scalar collision rate density}
   \begin{align}
    C(\pos) = \int_{4\pi} C(\pos,\dir) d \dir.
   \end{align}
   We denote the \emph{scalar flux} (fluence rate) $\phi(\pos)$, which is the angular integral of the \emph{vector-flux} (radiance) $\psi(\pos,\dir)$,
   \begin{equation}
    \phi(\pos) = \int_{4 \pi}\psi(\pos,\dir) d\dir.
   \end{equation}
   \removetwo{The angular flux describes particles in flight, whereas the collision rate density describes collisions and we strongly discourage attaching the label ``radiance'' to the latter~\cite{liemert2017radiative}.}

  \subsection{Classical Local Balance}

  Classical linear transport theory assumes that a particle at position $\pos$ traveling in direction $\dir$ traversing an incremental distance $ds$ has an incremental collision probability $dp$ given by
  \begin{equation}\label{eq:classicalassumption}
    dp = \Sigma_t(\pos,\dir) ds,
  \end{equation}
  which directly implies Beer-Lambert attenuation.  The \emph{transmittance}
  \begin{equation}\label{eq:beer}
    X(\pos,s,\dir) = \e^{-\int_0^s \Sigma_t(\pos + t \dir) dt}
  \end{equation}
  is the dimensionless probability of flying uncollided along a straight path between $\pos$ and $\pos + s \dir$ and is an exponential of the optical distance.  A closely related quantity is the distribution of free-path lengths $p(\pos,s,\dir)$, a probability density function (PDF) with units $m^{-1}$.  

  A \emph{free path} is the joint occurrence of two things: first flying uncollided along a path of length $s$ with probability $X(\pos,s,\dir)$, and then entering a collision in the next incremental path distance $ds$,
   \begin{equation}\label{eq:p}
   	 p(\pos,s,\dir) ds = X(\pos,s,\dir) \Sigma_t(\pos + s \dir) ds.
   \end{equation}
   In classical transport, the two components of the free path are independent: the collision process has no memory of how far the particle has flown.  This allows the inverse mean free path $\Sigma_t$ to locally convert between collision rates and fluxes.
   If the integral of the scalar flux $\phi(\pos)$ in some region $V$ is known and if $\Sigma_t(\pos)$ can be assumed constant in V, then the collision rate in $V$ can be found directly by multiplication by $\Sigma_t(\pos)$,
   \begin{equation}\label{eq:Cphiconversion}
     \int_{V} C(\pos) dV =  \Sigma_t(\pos) \int_{V} \phi(\pos) dV.
   \end{equation}
   Conversely, if $\Sigma_t(\pos) > 0$ then the flux integral can be found from the collision rate by division by $\Sigma_t(\pos)$.  This transformational power of $\Sigma_t$ is used in many other classical scenarios.  Monte Carlo methods estimating both reaction rates and flux integrals can use a single tally structure that subdivides phase space and decide after the simulation what quantities to derive from it.  Classical Green's functions and diffusion equations for flux are also convertible to those for collision rate by multiplication by $\Sigma_t$.

   \subsection{Local Imbalance in GRT}

   In GRT, the balance of collision rate and flux is nonlocal due to the short-term memory.  The cross sections are extended to depend on $s$~\cite{larsen11}, which gives an incremental collision probability following a collision of
   \begin{equation}\label{eq:GRTdp}
   	 dp = \Sigma_{tc}(\pos,\dir,s) ds
   \end{equation}
   and
   \begin{equation}
     dp = \Sigma_{tu}(\pos,\dir,s) ds
   \end{equation}
   if the previous scene interaction was a deterministic location, such as a boundary surface.
   This non-local collision probability breaks the classical balance of flux and collision rate\footnote{\newtwo{Note also that this same issue arises for the scalar quantities in anisotropic random media due to the dependence of $\Sigma_t(\dir)$ on direction.}}.  

\begin{figure}
      \centering
      \includegraphics[width=.5 \linewidth]{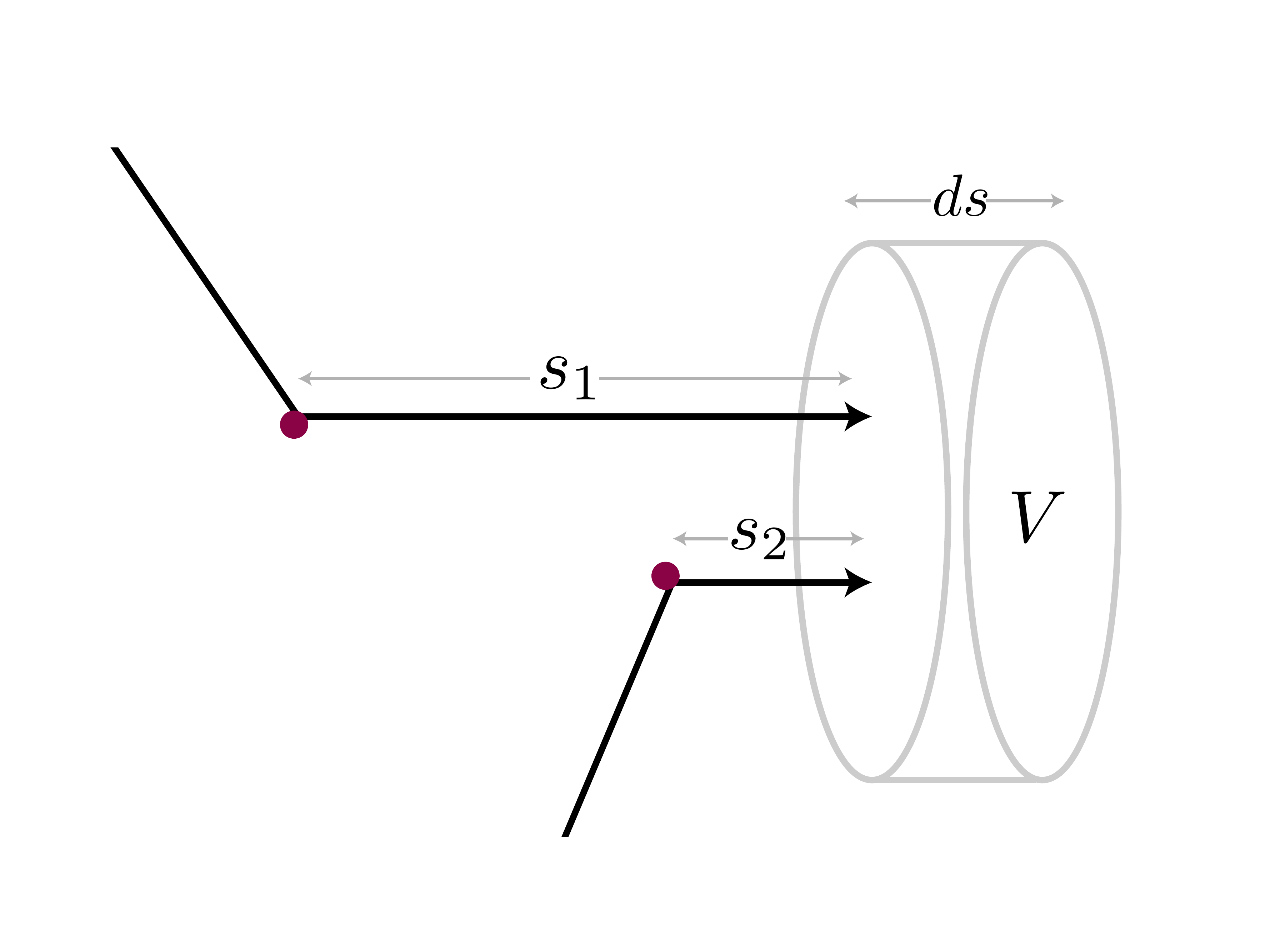}
      \caption{When the scatterers in a homogeneous random medium are spatially correlated, the probability of a collision \removetwo{is track-length dependent} \newtwo{includes a memory term}.  The macroscopic cross section $\Sigma_{tc}(s)$ depends on $s$, the distance since previous collision, and the probability of collision when traversing a path of incremental distance $ds$ is $\Sigma_{tc}(s) ds$.  The collision rate in $V$ is estimated by the traditional collision estimator, scoring the particle weight $W$ for each collision in the region.  However, a track-length estimator for the collision rate in $V$ must score $W \int_s^{s+ds} \Sigma_{tc}(s') ds'$ \removetwo{due to the track-length dependence on collision probability} \newtwo{to account for the collision probability changing with free-path length $s$}.  Conversely, for estimating the flux integral in $V$, the traditional track length estimator, scoring $W$-weighted track-lengths in $V$, maintains its familiar form, but the collision estimator must be modified to score $W / \Sigma_{tc}(s)$.  If both flux integral and collision rate are desired in $V$, two distinct tallies must be kept.}
      \label{fig-gen-sigmat} 
   \end{figure}

   Consider negatively correlated scattering centers in the context of Figure~\ref{fig-gen-sigmat}, where $\Sigma_{tc}(\pos,\dir,s)$ is low or possibly zero for small $s$ to prevent short distances between collisions. Having arrived at $V$, the two particles have different collision probabilities within $V$.  Particle $2$ may have less chance to collide in $V$ than particle $1$, or even see zero collision probability, because of the proximity of its previous collision to $V$.
The ratio of total track-lengths passing through $V$ to the total number of collisions created within $V$ will depend on the distribution of free-path lengths $s$ before crossing $V$ and, thus, on the state of the system outside of $V$.  Therefore, to estimate both flux-proportional and collision-rate-proportional reaction rates from a single tally in GRT would require tallying over $s$ and using
   \begin{equation}
    \int_{V} C(\pos) dV =  \int_{V} \int_0^\infty \Sigma_t(\pos,s) \phi(\pos,s) ds dV,
   \end{equation}
   where $\phi(\pos,s)$ is the spectral decomposition of the scalar flux over path-length parameter $s$,
   \begin{equation}
    \phi(\pos,s) = \int_{4\pi}\psi(\pos,\dir,s) d\dir.
   \end{equation}
   The $s$-spectral flux $\psi(\pos,\dir,s)$ decomposes the traditional angular flux over $s$ and is required to form a local integro-differential form of the transfer equation, requiring an extra integration over $s$~\cite{larsen11} and is related to the scalar flux by
   \begin{equation}
    \phi(\pos) = \int_{0}^\infty \phi(\pos,s) ds.
   \end{equation}
   Given, amongst other things, the goal of global fluence tallies in reactor design~\cite{martin2012challenges}, we would prefer to avoid increasing the dimensionality of tally structures.  We achieve this by generalizing Monte Carlo estimators for estimating both quantities, and by keeping separate tallies.

\section{Monte Carlo Estimators in GRT}\label{sec:MC}

  In this section we propose generalizations of collision and track-length estimators for time-independent monoenergetic linear transport problems in GRT.  We assume a phase space $\Gamma$ with positions $\pos$ and unit directions $\dir$ in Euclidean $\mathbb{R}^d$ space, $\P = (\pos,\dir) \in \Gamma$.
\removetwo{We also assume a piecewise homogeneous isotropic random media for brevity.
  The \emph{single-scattering albedo} $c$ and \emph{phase function} $P(\dir_i \rightarrow \dir_o)$ have the standard definitions and are assumed to be both independent of \emph{free-path length} $s$.  The rest of the scattering process is completely determined by specifying, for each piecewise-homogeneous volume region, one of six functions, $\Sigma_{tc}(s), \Sigma_{tu}(s), p_c(s), p_u(s), X_c(s), X_u(s)$, any one of which uniquely determine the other five.  These are summarized in Table~\ref{tab:notation} and we briefly review their definitions.  Additional information can be found in part 1 of this work~\cite{deon2018reciprocal}.  The definition of the transport problem is completed by also specifying internal sources, boundary sources and the reflectance and transmittance properties of any boundaries and imbedded objects.}

  \removetwo{In a correlated random medium the statistics of collision are different for paths that begin from a location chosen uniformly at random versus those that begin from a previous collision site~\cite{audic1993monte}.  We denote the free-path statistics between collisions, and their related quantities, using the label ``c'' for \emph{correlated-origin}: macroscopic cross-section $\Sigma_{tc}(s)$, free-path distribution $p_c(s)$, and transmittance $X_c(s)$.  They are related by Eq.~\ref{eq:p}.  Any emission in the scene (such as neutron emission in reactors or thermal emission in an atmosphere) that arises from the same correlated structure that governs the primary transport is accounted for by the \emph{correlated source} term $Q_c(\pos,\dir)$.}

  \removetwo{For all other free-paths, assumed to have origins that are spatially uncorrelated from scattering centers in the medium, the label ``u'' specifies the \emph{uncorrelated-origin} macroscopic cross-section $\Sigma_{tu}(s)$, free-path distribution $p_u(s)$, and transmittance $X_u(s)$.  A separate source term $Q_u(\pos,\dir)$ accounts for volume and boundary sources that find their first collisions using these statistics.}

  \subsection{Collision-Rate-Proportional Estimation}

    We first consider estimation of quantities $I$ that are linear functionals of the collision-rate density by specifying a measurement or ``detector'' function $g(\P$),
    \begin{equation}\label{eq:CI}
      I = \int_\Gamma g(\P) C(\P) d\P.
    \end{equation}
    \removetwo{Recall Eq.(\ref{eq:Cdefn}) for the definition of collision rate.}
    Monte Carlo estimators for $I$ are derived from the integral equation for $C$,
    \begin{multline}\label{eq:Cintegral}
      C(\pos,\dir) = \int_0^{s_b[\pos,-\dir]} \int_{4 \pi} K(\pos - s \dir,\dir' \rightarrow \pos,\dir) C(\pos - s \dir,\dir')  d\dir' ds \\ + S_0(\pos,\dir) + S_B(\pos,\dir)
    \end{multline}
    where $s_b[\pos,\dir]$ is a ray-tracing function that returns the distance $s_b$ from $\pos$ along $\dir$ to the nearest boundary.  
    The intercollision transport kernel $K$ combines the fraction of particles scattered, their change in direction and their next free path,
    \begin{equation}
      K(\pos',\dir' \rightarrow \pos,\dir) = c P(\dir' \rightarrow \dir;\pos') p_c(||\pos - \pos'||).
    \end{equation}
    To support reciprocal GRT, we have modified (\ref{eq:Cintegral}) from the previous form~\cite{larsen11} in two ways.  The density of initial collisions
    \begin{align*}
      S_0(\pos,\dir) &dV d\dir =  \\
      &\text{the rate at which particles scatter into $dV d\dir$ } \\
      &\text{about $(\pos,\dir)$ directly from sources $Q$}
    \end{align*}
    includes contributions from the new uncorrelated-origin source $Q_u$,
    \begin{equation}
      S_0(\pos,\dir) = 
        \int_0^{s_b[\pos,-\dir]} \left( Q_c(\pos - s \dir,\dir) p_c(s) + Q_u(\pos - s \dir,\dir) p_u(s) \right) ds.
    \end{equation}
    Here, $Q_u$ accounts for both sources at the boundary and internal sources emitting from medium-uncorrelated locations.  We also include an optional term $S_B$ that accounts for the first collisions that follow surface reflection and transmission events for problems with reflecting and refracting boundary conditions.  This requires knowing the angular flux leaving the boundary location $\pos - s_b[\pos,-\dir] \dir$ which is found by tracing along $-\dir$ from $\pos$.
    \begin{equation}
      S_B(\pos,\dir) = \psi(\pos-s_b[\pos,-\dir] \dir,\dir) p_u(s_b)
    \end{equation}
    The angular flux $\psi(\pos_b,\dir)$ leaving a boundary surface location $\pos_b$ in direction $\dir$ is found by integrating the angular flux arriving there $\psi_i(\pos_b,\dir')$ with the \new{b}idirectional surface reflection distribution function (BSDF)~\cite{pharr16} $f_s(\pos_b,\dir,\dir')$, which accounts for reflectance and transmittance,
    \begin{equation}
      \psi(\pos_b,\dir) = \int_{4 \pi} f_s(\pos_b,\dir,\dir') \psi_i(\pos_b,\dir') |\vec{n} \cdot \dir'| d\dir',
    \end{equation}
    where $\vec{n}$ is the surface normal.  The angular flux arriving at the surface \emph{from} direction $-\dir$ is
    \begin{multline}
      \psi_i(\pos_b,-\dir) = \psi(\pos_b - \dir s_b[\pos_b,-\dir],\dir) X_u(s_b[\pos_b,-\dir]) + \\ \int_0^{s_b[\pos_b,-\dir]} \int_{4 \pi} C(\pos - s \dir,\dir') c P(\dir' \rightarrow \dir) X_c(s) d\dir' ds,
    \end{multline}
    the sum of the flux scattered into $-\dir$ from collisions in the volume together with the uncorrelated transmittance of flux leaving the first surface found in that direction.

  \subsection{The Analog Walk for GRT}

    To build estimators for $I$, we begin with the analog walk for GRT, which is very similar to the classical case.  A random walk $w = (\pos_0,\dir_0),...,(\pos_k,\dir_k)$ begins by source sampling $(\pos_0,\dir_0,E)$ from the union of the sources, $Q_c$ and $Q_u$, in direct proportion to their relative \newtwo{powers}.  The initial vertex $(\pos_0,\dir_0)$ of $w$ is this sampled source location and direction.  The initial free-path length $s_0$ is sampled from either $p_c(s)$ (in the case $Q_c$ was sampled) or $p_u(s)$ (otherwise).  Additional vertices are added by repeating the following process.  Subsequent positions are determined via
    \begin{equation}
      \pos_{i+1} = \pos_i + \text{min}\left(s_i, \, s_b[\pos_i,\dir_i]\right) \dir_i.
    \end{equation} 
    Now, currently at position position $\pos_{i+1}$, either the phase function at the current volume interior point or analog\footnote{Analog sampling of BSDFs is uncommon in practice for surfaces that absorb energy.  To be precise, an analog sampling process for a BSDF returns weight $0$ (an absorption event) with a probability of $1 - \alpha(\dir_i)$ where $\alpha(\dir_i)$ is the albedo of the material for the current incoming direction $\dir_i$, and returns direction $\dir$ (with weight $1$) in proportion to $|\vec{n} \cdot \dir|f_s(\dir,\dir_i)$ otherwise.} sampling of the BSDF at the current boundary surface determine the next direction $\dir_{i+1}$.  If escape or absorption is sampled, the walk terminates.  Otherwise, a medium collision is followed with a new free-path length $s_{i+1}$ sampled from $p_c(s)$ and a surface interaction is followed with a sample from $p_u(s;\pos_{i+1})$ for the current volume region (which may have changed).

  \subsection{The Collision Estimator}

    Nothing about the inclusion of two sources, non-exponential free-path statistics or boundary reflections materially alters the proof~\cite{spanier69} that, for the analog random walk $w$,
    \begin{equation}\label{eq:collisionestimatoranalog}
      \sum_{i=1}^{k(w)} g(\pos_i,\dir_i)
    \end{equation}
    is an unbiased estimator for $I$.  With $g(\pos,\dir) = 1$ for all interior points of some region $V$ of the medium and $0$ elsewhere, Eq.(\ref{eq:collisionestimatoranalog}) is an unbiased estimate of the collision rate $\int_V C(\pos) dV$, in $V$.  By definition of the analog walk, collisions are created in the same proportion to the physical process they model, and so this is expected.  Note that, in our notation, vertices of $w$ corresponding to boundary surface interactions avoid contributing to the collision rate or measurement $I$ by defining $g$ to be $0$ exactly at the boundary.

    \removetwo{Determining the appropriate measurement function $g$ to use for a desired reaction rate of interest in a multi-species system is more complicated in GRT.  Recall that we have specified the transport behaviour directly with a free-path distribution or related quantity and that number densities of scattering particles have made no appearance (more on this in Section~\ref{sec:discussion}).  In some cases, due to the specific spatial correlations in the media, it may not be accurate to set $g$ to be a ratio of number densities or a ratio of classical cross-sections at the current energy $\Sigma_r(E) / \Sigma_t(E)$ to estimate the reaction rate caused by some isotope ``r''.  Consider, for example, a clumped microstructure where, in each clump, material of type A surrounds material of type B making collisions with matter of type B unlikely.  This would not be noticed in the sampling procedure proposed by Larsen and Vasques~\shortcite{larsen11} that estimates only $p_c(s)$ by tracing within explicit random realizations and accumulating free-path histograms, and so we recommend extending their sampling procedure to also verify that proportional reaction rates for collisions of interest are independent of free-path length and direction.}

    \subsubsection{Estimating Flux with the Collision Estimator}

      In classical linear transport, estimating flux using the collision estimator makes use of the classical balance of flux and collision rate.  For each collision, by Eq.(\ref{eq:classicalassumption}), there is, on average, a free-path of length $1 / \Sigma_t$ required to make that collision possible.  Thus, $g$ is set to $g(\pos,\dir) = 1 / \Sigma_t(\pos,\dir)$ in $V$.  This limits the measurement of flux to regions where collisions occur ($\Sigma_t(\pos) > 0$).  Where no collisions occur, alternative deterministic \remove{uncollided} calculations can often be applied~\cite{case53}.  As $\Sigma_t$ decreases, the performance of the collision estimator degrades, making the track-length estimator indispensable in optically-thin regions.

      In GRT, flux estimation with the collision estimator can have additional limitations.  We see immediately from Eq.(\ref{eq:GRTdp}) that the free-path lengths crossing $\pos$ required to create the known collision rate there \removetwo{include a track-length dependence}\newtwo{depend on $s$}, requiring a collision estimator with
      \begin{equation}\label{eq:collisionflux}
        g^\star(\pos,\dir,s) = \begin{cases}
          \frac{1}{\Sigma_{tc}(\pos,\dir,s)} & \text{for collisions with correlated origins} \\
          \frac{1}{\Sigma_{tu}(\pos,\dir,s)} & \text{otherwise}.
        \end{cases}
      \end{equation}
     Eq.(\ref{eq:collisionflux}) with  Eq.(\ref{eq:collisionestimatoranalog}) is an unbiased estimator for the flux, under certain restrictions, but we see that flux estimation in GRT is no longer a linear functional of the collision rate with a local kernel $g(\pos,\dir)$.  We have labeled this estimation with a star because $s$ is not in the phase space and therefore Eq.(\ref{eq:collisionflux}) is incompatible with Eq.(\ref{eq:CI}).  Ultimately, we define the class of flux-proportional measurements $I_\psi$ to be linear functionals of the spectral collision-rate density,
     \begin{equation}
      I_\psi = \int_{\Gamma^\star} g^\star(\P^\star) C(\P^\star) d\P^\star
     \end{equation}
     where we have extended the phase space $\P^\star = (\pos,\dir,s) \in \Gamma^\star$ to include free-path parameter $s$. The \emph{spectral collision-rate density} is
     \begin{align}
      C(\pos,\dir,s) &dV d\dir ds \\ = \, &\text{rate at which particles in $dV d\dir ds$ } \nonumber \\
                               &\text{about $(\pos,\dir,s)$ enter collisions } \nonumber
     \end{align}
     providing a decomposition of the collision density over the distance $s$ since previous scene interaction.
     The analog random walk previously defined can also be used to estimate measurements of class $I_\psi$ because the path-length $s$ is known at each step in the walk generation, so Eq.(\ref{eq:collisionflux}) with Eq.(\ref{eq:collisionestimatoranalog}) is a practical estimator in some cases.

     Similar to the classical case, the performance of Eq.(\ref{eq:collisionflux}) will be limited by the lowest value of $\Sigma_{tc}(s)$ over $s$.  Note that this is no longer a local limitation applying only to optically thin regions, but to the entire piecewise homogeneous volume element.  Furthermore, \newtwo{if there exists a range of finite width $0 < a < s < b$ with $\Sigma_{tc}(s)$ identically zero on this range while also $\Sigma_{tc}(s) > 0$ for some portion of $s > b$}, then the free path distribution has a gap where no collisions ever occur, but where flux arises from traversing that gap and colliding later with non-zero probability.  \new{If so, then} this collision estimator is incompatible with flux estimation \emph{everywhere}.

     \newtwo{\emph{Example 1:} Consider negatively correlated scattering centers in a 3D slab with uniform plane-parallel illumination arriving normally on one side (Figure~\ref{fig-pcpupretty}).  Figure~\ref{fig-example1} illustrates the inability of the collision estimator $1 / \Sigma_{tc}(s)$ to estimate flux inside a unit thickness slab with intercollision macroscopic cross-section
     \begin{equation}\label{eq:bluesigma}
      \Sigma_{tc}(s) = \begin{cases}
                        0 & s < \hat{s} \\
                        \frac{1}{\ell - \hat{s}} & \text{else}
                        \end{cases}
     \end{equation}
     due to the optically empty gap $0 < s < \hat{s}$.  In this example, we consider conservative ($c = 1$) isotropic scattering, with $\ell = 0.333$ and minimum intercollision free-path lengths of $\hat{s} = 0.065$.
      \begin{figure}
      \centering
      \includegraphics[width=.8 \linewidth]{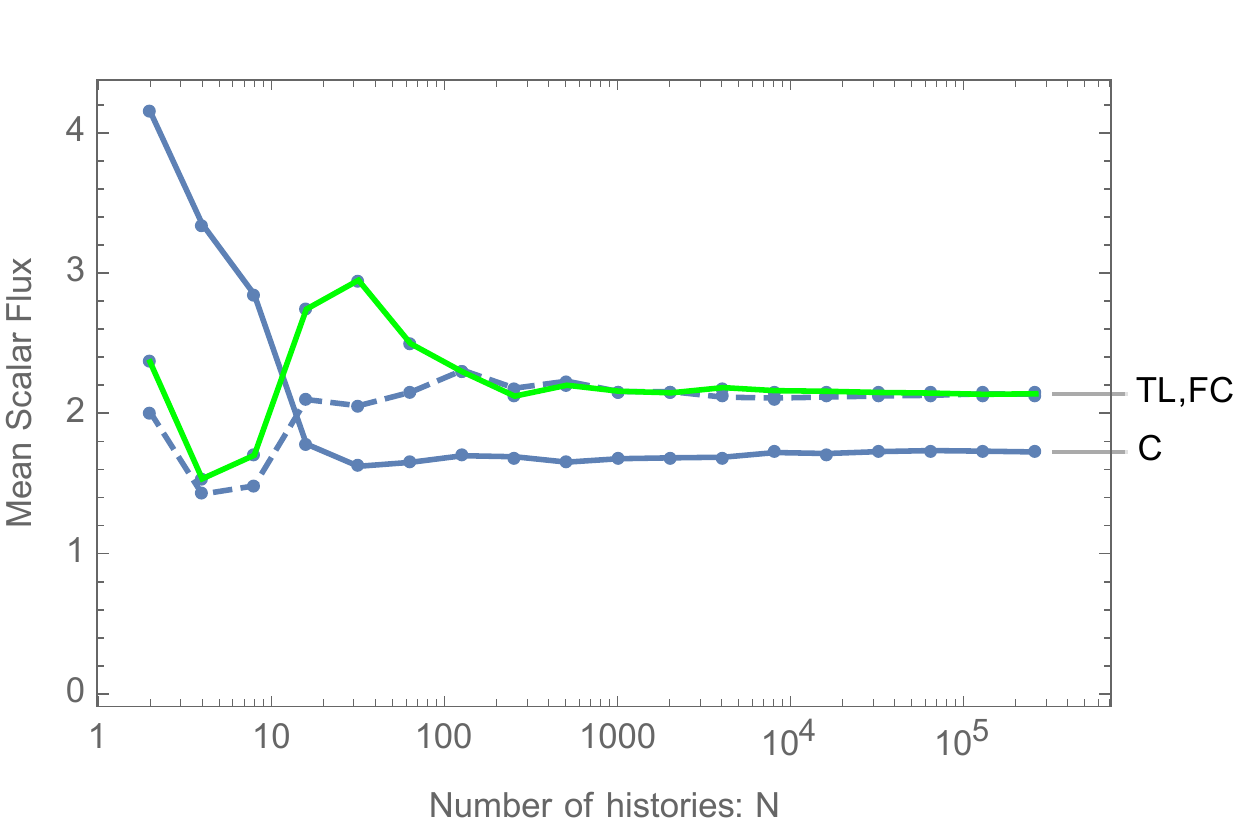}
      \caption{\newtwo{Comparison of Monte Carlo estimation of the total scalar flux in a purely-scattering 3D slab with negatively-correlated scatterers.  The classical collision estimator (C) underestimates the flux because the macroscopic cross-section (Equation~\ref{eq:bluesigma}) has an optically-empty gap between collisions.  A modified collision estimator with fictitious scattering (FC,dashed) or a track-length estimator (TL,green) is required to accurately estimate the flux.}}
      \label{fig-example1} 
    \end{figure}
    }

    \subsubsection{Fictitious Scattering}\label{sec:mentalmodel}

      Some of the limitations of the collision estimator for flux in GRT can be mitigated by introducing fictitious matter along the free-path sampling portion of the walk such that collision with the fictitious matter ultimately causes the particle to scatter forward with no loss, as if the matter wasn't there.  This applies the Woodcock/delta-tracking method used for heterogeneous volumes, but in a homogeneous setting, analogous to the class of estimators introduced by Spanier~\cite{spanier66,lux78,lux78b}.  Increased density of collision in ``optically thin'' portions of the free path lower the variance of the estimator in those regions.  The trade-off is additional cost of computation time to perform the walk.

      Intuition for this generalization is found from conceptualizing correlated random media as a classical heterogeneous medium that changes each time a collision occurs due to its short term memory.  The last sampled collision in the medium collapses the set of all possible random realizations down to only those where such a scattering event could have occurred, applying $\Sigma_{tc}(r)$ radially around the point of the last known collision.  Using Woodcock/delta tracking to form random walks within this collapsed medium in the presence of additional fictitious density, keeping the real density field $\Sigma_{tc}(r)$ fixed until a real collision is sampled, extends the collision estimator and, in the limit of infinite fictitious density everywhere, yields the track-length estimators for GRT.  In the case of an uncorrelated free path, such as entry at a boundary, the medium is initialized to equilibrium with all possible random realizations in equal probability.  As this path extends into the medium, the memory of finding no collisions up to distance $s$ continuously collapses the nature of the medium via $\Sigma_{tu}(s)$ (which is why $\Sigma_{tu}(s)$ \new{is not a constant}) until a collision occurs and then the medium snaps to $\Sigma_{tc}(r)$.

      The collision estimator for GRT is extended with fictitious scattering by choosing $\Sigma^\star(s)$ such that $\Sigma^\star(s) \ge \Sigma_{t\_}(s)$ for all $s > 0$.  Both correlated- and uncorrelated-origin free paths may independently perform the generalizations in this section, choosing the same or different $\Sigma^\star(s)$.  We describe simply $\Sigma_{t\_}(s)$ for this derivation to mean either.

      Having chosen $\Sigma^\star$, free path lengths for the walk $w$ are modified to sample from
      \begin{equation}
        p^\star(s) = \Sigma^\star(s) \e^{-\int_0^s \Sigma^\star(s')ds'}.
      \end{equation}
      At each medium collision of the walk, the collision type (real vs fictitious) is decided randomly with probability $p_\delta$ for fictitious collisions
      \begin{equation}
        p_\delta(s) = \frac{\Sigma^\star(s) - \Sigma_{t\_}(s)}{\Sigma^\star(s)}.
      \end{equation}
      Both real and fictitious collisions add vertices to the walk $w$.  Source sampling and boundary collisions are treated identically to the standard case.  Real collisions determine $\dir_{i+1}$ from the phase function with probability $c$, resetting $s$ to $0$, and terminate the walk with absorption otherwise.  Fictitious collisions continue walking along $\dir_{i+1} = \dir_i$ with probability $1$ but $s$ is \emph{not reset to $0$}---a critically important difference for applying this method to GRT.  This requires sampling tails of $p^\star(s)$.  If the current distance to the previous real collision is $s_r$, then we must sample the next free-path length $s$ from,
      \begin{equation}
        p(s) = \frac{p^\star(s+s_r)}{\int_{s_r}^\infty p^\star(s') ds' }.
      \end{equation}
      If $\Sigma^\star$ is a constant that bounds $\Sigma_{t\_}(s)$ everywhere, then the tails are exponential and all free-path lengths can be sampled using $s = -\frac{1}{\Sigma^\star}\log \xi$, where $\xi \in (0,1)$ is a uniform random number.  However, a majorant of $\Sigma_{t\_}(s)$ might not exist or use of one might not be desired (due to inefficiency), requiring a more general $\Sigma^\star(s)$.  We discuss tail sampling further in Section~\ref{sec:tails}.  

      Collisions now occur $1 / (1-p_\delta(s))$ more frequently than in the analog walk.  To form an unbiased estimator, the fictitious collision estimator must weight the scores at each collision vertex by the inverse of this proportionality,
      \begin{equation}\label{eq:collisionestimatordelta}
        \sum_{i=1}^{k(w)} g^\star(\pos_i,\dir_i) (1 - p_\delta(s_{i-1})) = \sum_{i=1}^{k(w)} g^\star(\pos_i,\dir_i) \frac{\Sigma_t(s)}{\Sigma^\star(s)}.
      \end{equation}
      This estimator can fill empty or low portions of the correlated free-path distribution to permit flux estimation in scenarios where it would otherwise fail or perform poorly.  In the limit that $\Sigma^\star(s) \rightarrow \infty$ for all $s$, we form estimators of the track-length type that score continuously along segments of the walk.

      \newtwo{\emph{Example 1 (continued):} Returning to Example 1, defined above, we form a new collision estimator  that fills the optically-empty portion of the intercollision free paths with fictitious matter using $\Sigma^\star = 1 / (\ell - \hat{s})$, leaving the collision estimator for the uncorrelated entry step unchanged ($1 / \Sigma_{tu}(s)$).  Figure~\ref{fig-example1} shows the new collision estimator converging to the exact solution, which we verify using a track-length estimator, described next.  For $N = 10^8$ histories, the fictitious collision estimator estimated a mean of $2.1658$ in $34.4$s with a variance of $5.07$.  The track-length estimator estimated a mean of $2.1340$ in $27.4$s with a variance of $4.36$.}

   \subsection{The Track-length Estimator for GRT}\label{sec:tracklength}

     Taking the limit in Eq.(\ref{eq:collisionestimatordelta}) that, for all $s > 0$, $\Sigma^\star(s) \rightarrow \infty$, following the same arguments as the classical derivation~\cite{spanier66}, we find the track-length estimator for GRT,
     \begin{equation}
       \sum_{i=0}^{k(w)-1} \int_0^{||x_{i+1}-x_i||} g^\star(\pos_i + s \dir_i,\dir_i,s) \Sigma_{t\_(i)}(s) ds.
     \end{equation}
     Here, the notation $\Sigma_{t\_(i)}$ means
     \begin{equation}
        \Sigma_{t\_(i)} = \begin{cases}
          \Sigma_{tc} & \text{if $\pos_i$ is a medium-correlated origin} \\
          \Sigma_{tu} & \text{if $\pos_i$ is a medium-uncorrelated origin}.
        \end{cases}
      \end{equation}

     \subsubsection{The Track-length Estimator for Flux}

       Noting that the collision estimator in Eq.(\ref{eq:collisionflux}) for flux in some region $V$  scores the inverse of the path-length dependent macroscopic cross-section in that region and $0$ elsewhere, we see that this term cancels in the track-length estimator.  Hence, the track-length estimator reduces to scoring the total length of all tracks passing through $V$, and so is, in some sense, \emph{the} track-length estimator in that it scores the length of the track (as opposed to scoring along it).  Just as it was no surprise that the collision estimator for collision rate required no change in GRT, here too we expect this result, especially in light of the International Commission on Radiation Units and Measurements definition of fluence~\cite{lux91}.

     \subsubsection{The Track-length Estimator for Collision Rate}\label{sec:TLintegrals}

      Noting that the collision estimator for collision rate in some region $V$ scores $1$ in that region and $0$ elsewhere, we see that the track-length estimator in GRT for collision rate scores,
      \begin{equation}
        \int_{s_1}^{s_2} \Sigma_{t\_}(s) ds,
      \end{equation}
      the integral of $\Sigma_{t\_}(s)$ for any portion of a track that lies within $V$, where $
      \{s_1,s_2\}$ indexes that track subsegment.  In some scenarios, such as when the free-path distribution has been determined by the histogram sampling approach~\cite{audic1993monte,moon07,larsen11}, this estimator would be the least attractive of the four, in needing to compute unique integrals per score.  These integrals may have no simple closed form or may be inconvenient to derive from a discrete histogram.  \new{However,} for many parametric analytic correlated forms of transport, this limitation is not much of a concern.  

      Later in the paper we consider two specific non-exponential free-path distributions.  The Gamma-2 distribution $p_c(s) = \e^{-s} s$ has $\Sigma_{tc}(s) = \frac{s}{1+s}$ and therefore scores
      \begin{equation}
        \int_{s_1}^{s_2} \Sigma_{tc}(s) ds = s_2 - s_1 - \log \left( \frac{1+s_2}{1+s_1} \right)
      \end{equation}
      for correlated-origin tracks and
      \begin{equation}
        \int_{s_1}^{s_2} \Sigma_{tu}(s) ds = s_2 - s_1 - \log \left( \frac{2+s_2}{2+s_1} \right).
      \end{equation}
      for uncorrelated-origin tracks.

      We also consider a power-law attenuation law~\cite{davis14}, adopting the reciprocal imbedding proposed in~\cite{deon2018reciprocal,jarabo18}.  The correlated free path distribution is then
      \begin{equation}
        p_c(s) = a (a+1) \ell (a \ell)^a (a \ell+s)^{-a-2}
      \end{equation}
      with correlated mean free path length $\s = \ell$ and shape parameter $a > 1$. Exponential attenuation is recovered in the limit $a \rightarrow \infty$.  The track-length integrals for power-law transport are
      \begin{align}
        &\int_{s_1}^{s_2} \Sigma_{tc}(s) ds = (a+1) \log \left(\frac{a \ell+s_2}{a \ell+s_1}\right) \\
        &\int_{s_1}^{s_2} \Sigma_{tu}(s) ds = a \log \left(\frac{a \ell+s_2}{a \ell+s_1}\right) 
      \end{align}
      This power-law free-path distribution was derived from a continuum model of stochastic density with Gamma-distributed noise of parameter $a$~\cite{davis2011radiation}.  Using the inverse Laplace transform of $p_c(s)$ (with $\s = \ell = 1$) we find that this free-path distribution is a Gamma-distributed superposition of positive exponential free-path distributions,
      \begin{equation}
        p_c(s) = \int_0^\infty   \frac{a^a e^{-a \Sigma_t} \Sigma_t^a}{\Gamma (a)}  \Sigma_t \e^{-\Sigma_t s} d\Sigma_t.
      \end{equation}
      One interpretation of this is that a free path in this family of correlated random media with mean cross-section $\hat{\Sigma_t} = 1$ is equivalent to choosing a random classical cross-section $\Sigma_t > 0$ from the normalized gamma distribution
      \begin{equation}
        \frac{a^a e^{-a \Sigma_t} \Sigma_t^a}{\Gamma (a)}
      \end{equation}
      and keeping this cross section fixed until the next collision.

      \new{To test the new Monte Carlo estimators, we derive benchmark solutions for infinite homogeneous media in the next section.  Additional discussion of implicit capture, zero-variance guiding and the duality of two classes of estimators is included in Appendix~\ref{appendix:MC}.}

\section{Green's Functions for the Isotropic Point Source}\label{sec:greens}

  \new{We now turn to the derivation of} exact solutions for the monoenergetic time-independent scalar collision rate density $C(\pos)$ and scalar flux $\phi(\pos)$ about an isotropically-emitting point source in an infinite isotropically-scattering medium.  In the case of emission that is spatially correlated to the scattering centers in the medium, we reproduce the derivations of a previous work~\cite{deon14} for the \emph{correlated collision rate Green's function} $C_c(\pos)$ and \emph{correlated scalar flux Green's function} $\phi_c(\pos)$.  Two new solutions, $C_u(\pos)$ and $\phi_u(\pos)$, for the case of uncorrelated emission, are derived to form a complete set of infinite medium Green's functions for reciprocal generalized linear transport. 

  Consider a unit \newtwo{power} isotropic point source at the origin of the infinite homogeneous medium with single-scattering albedo $0 < c < 1$ and isotropic scattering.  \new{The scalar collision-rate density $C(\pos)$ at position $\pos \in \mathbb{R}^d$ is the solution of the generalized Peierls integral equation~\cite{grosjean51,larsen11}
  \begin{equation}\label{eq:Peirels}
    C(\pos) = C_0(\pos) + c \int_{\mathbb{R}^d} C(\pos') \frac{p_c(||\pos-\pos'||)}{\Omega_d(||\pos-\pos'||)} d\pos',
  \end{equation}
  where the surface area of the sphere of radius $r$ in $\mathbb{R}^d$ is written
  \begin{equation}
    \Omega_d(r) = \frac{d \pi ^{d/2} r^{d-1}}{\Gamma \left(\frac{d}{2}+1\right)},
  \end{equation}
  and $C_0(\pos)$ is the collision rate density of first collisions in the system arising directly from two isotropically-emitting sources,
  \begin{equation}
    C_0(\pos) = \int_{\mathbb{R}^d} Q_c(\pos') \frac{p_c(||\pos-\pos'||)}{\Omega_d(||\pos-\pos'||)} d\pos' + \int_{\mathbb{R}^d} Q_u(\pos') \frac{p_u(||\pos-\pos'||)}{\Omega_d(||\pos-\pos'||)} d\pos'.
  \end{equation}
  For the case of a single point source, exactly one of $Q_c(\pos)$ and $Q_u(\pos)$ is a Dirac delta $\delta(\pos)$ at the origin and the other is $0$.}

  \new{Equation~\ref{eq:Peirels} differs from the case of classical radiative transfer by permitting arbitrary free-path length statistics between collisions via $p_c(s)$, and reduces to the familiar form when all free-path length statistics are exponential, $p_c(s) = p_u(s) = \Sigma_t \exp \left[ -s \, \Sigma_t \right]$ (the unique scenario where $Q_c$ and $Q_u$ become equivalent).  As in the classical case, the primary challenge is solving for $C(\pos$).
  Once $C(\pos)$ is known, the scalar flux $\phi(\pos)$ follows directly from spatial integration with a transmittance kernel, described below.} 

  A Neumann series solution of Equation~\ref{eq:Peirels} can be formed by repeated convolution of uncollided propagators, interleaved with absorption.  We employ a Fourier transform approach~\cite{grosjean51,zoia11e}, where the required convolutions are easily formed.  The radial symmetry of the emission produces solutions depending on a single spatial coordinate $r$, the distance from the point source, so this problem is typically termed a 1D transport problem, but it describes a symmetric transport process in $\mathbb{R}^d$.

  We will make use of the following forward and inverse spherical Fourier transforms of a radially-symmetric function $f(r)$ (with radius $r \ge 0$) in $\mathbb{R}^d$~\cite{dutka85}
	\begin{align}\label{eq:sphericalFourier}
		&\bar{f}(z) = \F\{f(r)\} = z^{1-d/2} (2\pi)^{d/2} \int_0^\infty r^{d/2} J_{d/2-1}(r z) f(r) dr \\
		&f(r) = \Finv\{\bar{f}(z)\} = \frac{r^{1-d/2}}{(2\pi)^{d/2}}  \int_0^\infty z^{d/2} J_{d/2-1}(r z) \bar{f}(z) dz
	\end{align}
	where $z$ is the transformed coordinate relating to $r$ and $J_k$ is the modified Bessel function of the first kind.

  \subsection{Collision Rate Density}

    We now introduce the \emph{correlated-origin propagator}
  	\begin{equation}
  		\zeta_c(r) = \frac{p_c(r)}{\Omega_d(r)}
  	\end{equation}
  	and the \emph{uncorrelated-origin propagator}
  	\begin{equation}
  		\zeta_u(r) = \frac{p_u(r)}{\Omega_d(r)}
  	\end{equation}
    where the probability of finding a particle entering a collision within $dr$ of distance $r$ from its previous medium event is $\zeta_c(r) dr$ if the previous event was a medium collision or birth from a source correlated in location to the scatterers, and $\zeta_u(r) dr$ otherwise.
  	These distributions describe only spatial translations between collision events and are energy-conserving,
  	\begin{equation}
  		\int_0^\infty \Omega_d(r) \zeta_c(r) dr = \int_0^\infty \Omega_d(r) \zeta_u(r) dr = 1.
  	\end{equation}
  	Absorption is assumed to occur at the scattering centers and the single-scattering albedo $c$ is assumed independent of the track length $s$ between collisions.

   Combining the propagators and $c$, the Neumann series solution of Equation~\ref{eq:Peirels} is readily formed as a geometric series of convolutions.  For example, the density of particles entering their second collision after correlated birth is given by the absorption-weighted convolution $ c \, \zeta_c(r) \star \zeta_c(r) $, the third collision density by $ c^2 \, \zeta_c(r) \star \zeta_c(r) \star \zeta_c(r) $, and so on.  We denote the Fourier transforms of the propagators
  	\begin{align}\label{eq:Fourierprop}
  		& \F\{\zeta_c(r)\} = \bar{\zeta}_c(z), \\
  		& \F\{\zeta_u(r)\} = \bar{\zeta}_u(z)
  	\end{align}
    and solve for the various Green's functions in the frequency domain.  Exact or numerical Fourier inversion is used in numerical application of these results, which we discuss at the end of this section.

    \subsubsection{Correlated Emission (Collision rate)}

    	In the frequency domain, the initial distribution of collisions for correlated emission is given directly by $\bar{\zeta}_c(z)$, with subsequent collision orders found by repeated application of the kernel $c \, \bar{\zeta}_c(z)$ that includes both absorption and further propagation.  Thus, the total collision rate density including all orders of scattering is
	    \begin{equation}\label{eq:Cc}
	  		C_c(r) = \Finv\left\{ \frac{\bar{\zeta}_c(z)}{1 - c \bar{\zeta}_c(z)} \right\}.
	  	\end{equation}
	  	Separating out the initial collision density from the multiply-collided portion yields,
	  	\begin{equation}\label{eq:Cc2}
	  		C_c(r) = \frac{p_c(r)}{\Omega_d(r)} + \Finv\left\{ \frac{c \, ( \bar{\zeta}_c(z))^2}{1 - c \bar{\zeta}_c(z)} \right\},
	  	\end{equation}
	  	which is recommended when numerically inverting the Fourier solution, and also used later for the derivation of non-classical diffusion approximations.

	 \subsubsection{Uncorrelated Emission (Collision rate)}

	  	In the case of uncorrelated emission, the initial propagation is given by $\zeta_u(r)$, and all subsequent propagations are given by the kernel $c \, \bar{\zeta}_c(z)$, so the uncorrelated Green's functions for collision rate density are related to the correlated case by a factor of $\frac{\bar{\zeta}_u(z)}{\bar{\zeta}_c(z)}$, giving
	    \begin{equation}\label{eq:Cu}
	  		C_u(r) = \Finv\left\{ \frac{\bar{\zeta}_u(z)}{1 - c \bar{\zeta}_c(z)} \right\}
	  	\end{equation}
	  	and
	  	\begin{equation}\label{eq:Cu2}
	  		C_u(r) = \frac{p_u(r)}{\Omega_d(r)} + \Finv\left\{ \frac{c \,  \bar{\zeta}_c(z) \, \bar{\zeta}_u(z)}{1 - c \bar{\zeta}_c(z)} \right\}.
	  	\end{equation}

  \subsection{Scalar Flux \newtwo{/ Fluence}}

    The scalar flux about the isotropic point source is given by the sum of the uncollided flux leaving the collisions and the uncollided flux leaving the point source directly.  A frequency domain construction requires use of the \emph{stretched transmittance for correlated emission}
    \begin{equation}\label{eq:Xc}
    	\chi_c(r) = \frac{X_c(r)}{\Omega_d(r)},
    \end{equation}
    a related distribution for uncorrelated emission
    \begin{equation}\label{eq:Xu}
    	\chi_u(r) = \frac{X_u(r)}{\Omega_d(r)}
    \end{equation}
    and their Fourier transforms,
    \begin{align}
  		& \F\{\chi_c(r)\} = \bar{\chi}_c(z), \\
  		& \F\{\chi_u(r)\} = \bar{\chi}_u(z).
  	\end{align}

    \subsubsection{Correlated Emission (Flux)}

        In the case of correlated emission, the uncollided flux directly from the point source is given by $\chi_c(r)$ and the flux from the collisions is $\bar{\chi}_c(z) \, c \, \bar{C}_c(z)$ giving a total of

	    \begin{equation}
	  		\phi_c(r) = \Finv\left\{ \bar{\chi}_c(z) + \bar{\chi}_c(z) \, c \, \frac{\bar{\zeta}_c(z)}{1 - c \bar{\zeta}_c(z)} \right\} = \Finv\left\{ \frac{\bar{\chi}_c(z)}{1 - c \bar{\zeta}_c(z)} \right\}
	  	\end{equation}
	  	or, keeping the uncollided flux separate,
	  	\begin{equation}
	  		\phi_c(r) = \frac{X_c(r)}{\Omega_d(r)} + \Finv\left\{ \frac{c \, \bar{\chi}_c(z) \, \bar{\zeta}_c(z) }{1 - c \bar{\zeta}_c(z)} \right\}.
	  	\end{equation}

      Given the proportionality between uncorrelated free-path distribution and attenuation, we also have that the Green's functions $C_u(r)$ and $\phi_c(r)$ are proportional
      \begin{equation}\label{eq:phiCbalance}
        \phi_c(r) = \s C_u(r).
      \end{equation}

	\subsubsection{Uncorrelated Emission (Flux)}

	  For the case of uncorrelated emission, the uncollided flux directly from the point source is given by $\chi_u(r)$ and the flux from the collisions is $\bar{\chi}_c(z) \, c \, \bar{C}_u(z)$ giving a total of

	    \begin{equation}\label{eq:phiu}
	  		\phi_u(r) = \Finv\left\{ \bar{\chi}_u(z) + \bar{\chi}_c(z) \, c \, \frac{\bar{\zeta}_u(z)}{1 - c \bar{\zeta}_c(z)} \right\}
	  	\end{equation}
      or
      \begin{equation}\label{eq:phiu2}
        \phi_u(r) = \frac{X_u(r)}{\Omega_d(r)} + \Finv\left\{ \frac{c \, \bar{\chi}_c(z) \, \bar{\zeta}_u(z)}{1 - c \bar{\zeta}_c(z)} \right\}.
      \end{equation}

  \subsection{Numerical Validation}\label{sec:validation}

    We compared analytic and semi-analytic numerical predictions of the Green's functions derived in this section to Monte Carlo estimations of the corresponding collision rate and flux quantities and noted no difference apart from Monte Carlo noise.  A variety of absorption levels ($c \in \{ 0.01, 0.1, 0.3, 0.5, 0.7, 0.8, 0.9, 0.95, 0.99 \}$) were tested for transport in three dimensions with a correlated free-path distribution given by
    \begin{equation}\label{eq:freegamma}
      p_c(s) = \e^{-s} s
    \end{equation}
    with intercollision mean free path $\s = 2$.
    For both \new{collision} rate and flux integral quantities, we tested collision estimators for the point source problem and tested track-length estimators for an equivalent plane-source problem.  Our findings are briefly summarized below.  For additional validation of Eqs.~(\ref{eq:Cc}) or~(\ref{eq:Cc2}) for the case of correlated emission with other free-path distributions and general spatial dimensions ($d \in \{ 2, 3, 4\}$) see d'Eon~\shortcite{deon14}.

    \subsubsection{Point-Source Validation}

\begin{figure}
      \centering
      \hspace*{-1.6cm}
      \includegraphics[width=1.2\linewidth]{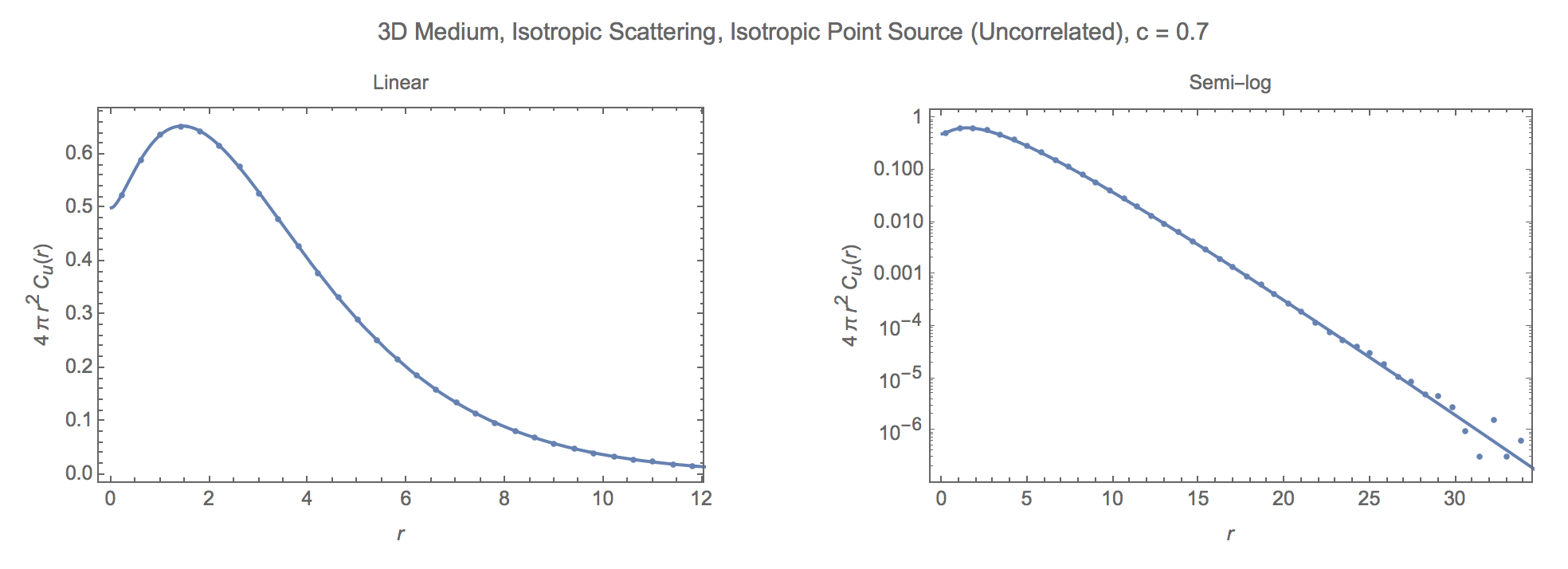}
      \caption{\label{fig:Cu}Collision estimation of collision rate density $C_u(x)$ at a distance $r$ from an isotropic uncorrelated point source in infinite 3D media with isotropic scattering and correlated free-path distribution $p_c(s) = s \e^{-s}$.  Monte Carlo (dots) vs analytic (continuous, Equation~\ref{eq:gamma2Cu}).} 
    \end{figure}

    The Gamma free-path distribution in Eq.~\ref{eq:freegamma}
     leads to uncorrelated free-paths~\cite{deon2018reciprocal}
    \begin{equation}\label{eq:freegammau}
      p_u(s) = \frac{1}{2} \e^{-s} (1 + s)
    \end{equation}
    and macroscopic cross-sections given by
    \begin{align}
      &\Sigma_{tc}(s) = \frac{s}{1+s}, \\
      &\Sigma_{tu}(s) = \frac{1+s}{2+s}.
    \end{align}
    The free propagators in the frequency domain, found by combining Eqs~\ref{eq:freegamma} and~\ref{eq:freegammau} with Eqs~\ref{eq:Fourierprop}, are
    \begin{align}
      &\bar{\zeta}_c(z) = \frac{1}{1+z^2}, \\
      &\bar{\zeta}_u(z) = \frac{1}{2} \left( \frac{1}{1+z^2}+\frac{\tan ^{-1}(z)}{z} \right).
    \end{align}
    From Eq.~\ref{eq:Cc}, the correlated collision rate density in the frequency domain is
    \begin{equation}\label{eq:testCc}
      \bar{C_c}(z) = \frac{1}{1-c+z^2},
    \end{equation}
    which inverts into the scalar diffusion point-source Green's function for the 3D medium,
    \begin{equation}
      C_c(r) =  \Finvthree\left\{ \frac{1}{1-c+z^2} \right\} = \frac{e^{-r \sqrt{1-c} }}{4 \pi  r}.
    \end{equation}
    For the case of collision rate due to uncorrelated emission, using Eq.~\ref{eq:Cu2}, we find\footnote{This integral can be solved by solving an equivalent problem in plane geometry and using the plane to point transform.  See Appendix~\ref{appendix:gamma2flux}.}
    \begin{equation}\label{eq:gamma2Cu}
      C_u(r) = \frac{e^{-r} (1+r)}{8 \pi  r^2}+\frac{c}{4 \pi^2 r}\int_0^{\infty } \frac{ \left(\frac{z}{1+z^2}+\tan
   ^{-1}(z)\right) }{  \left(1-c+z^2\right)} \sin (r z) \, dz,
    \end{equation}
    which we compare in Figure~\ref{fig:Cu} to Monte Carlo estimation using the analog collision estimator.

\begin{figure}
      \centering
      \subfigure[Correlated emission (Equation~\ref{eq:gammaphic})]{\hspace*{-1.6cm} \includegraphics[width=1.2\linewidth]{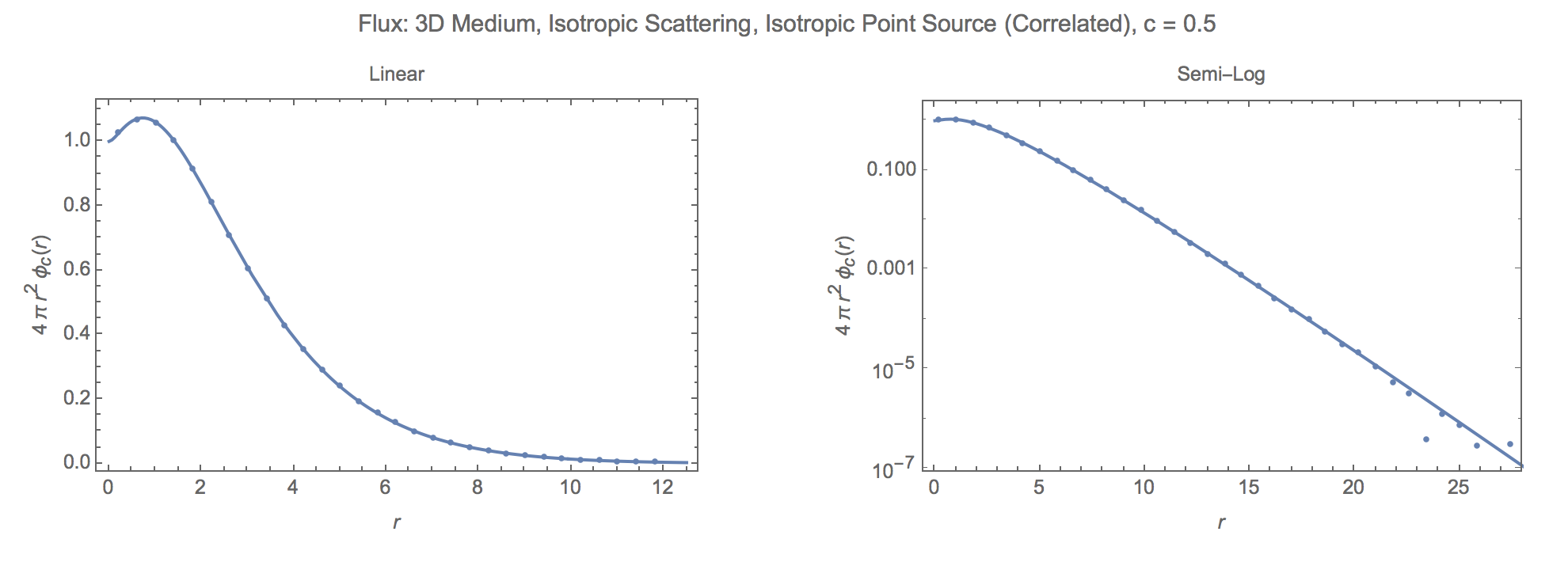}}
      \subfigure[Uncorrelated emission (Equation~\ref{eq:gammaphiu})]{\hspace*{-1.6cm} \includegraphics[width=1.2\linewidth]{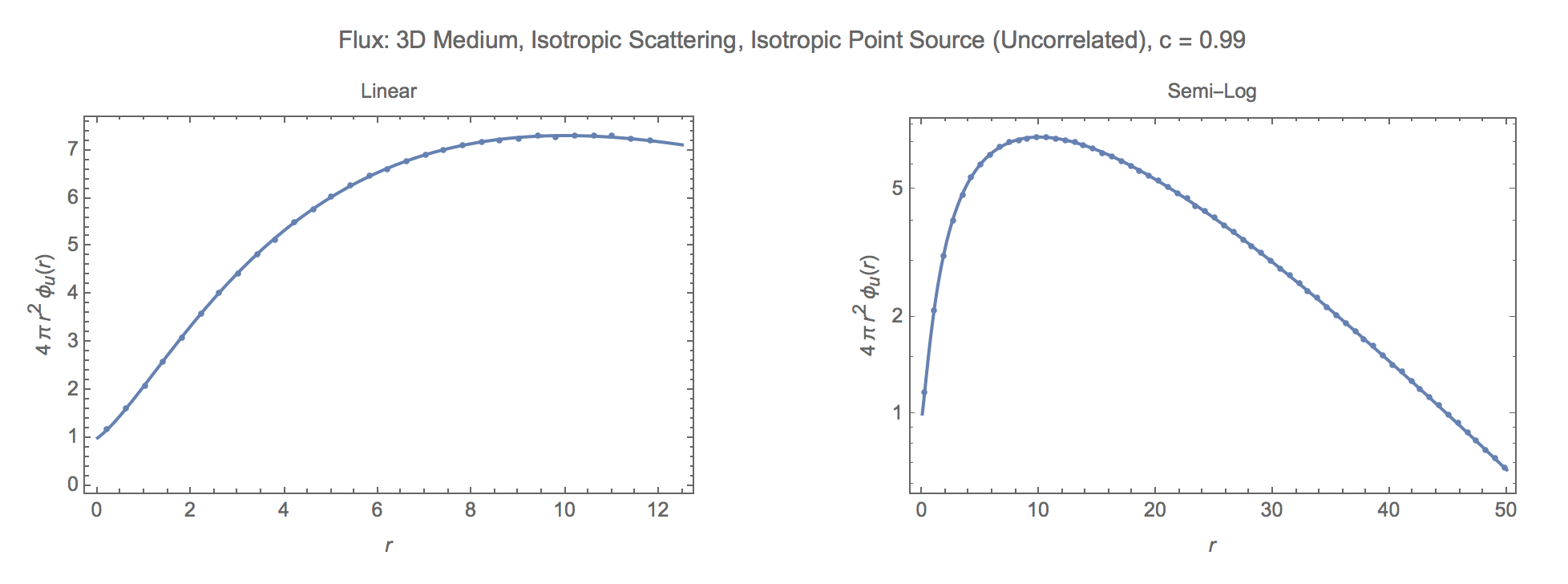}}
      \caption{Flux about an isotropic point source in a 3D medium.  Monte Carlo collision estimator (dots) vs analytic (continuous).}
      \label{fig-collisionflux} 
    \end{figure}

    For the stretched transmittances, we find (Eqs.~\ref{eq:Xc} and ~\ref{eq:Xu})
    \begin{align}
      &\bar{\chi}_c(z) = \frac{1}{z^2+1}+\frac{\tan ^{-1}(z)}{z}, \\
      &\bar{\chi}_u(z) = \frac{1}{2 z^2+2}+\frac{\tan ^{-1}(z)}{z},
    \end{align}
    producing the scalar flux Green's function about a correlated point source,
    \begin{equation}\label{eq:gammaphic}
      \phi_c(r) = 2 C_u(r)
    \end{equation}
    and the scalar flux Green's function about an uncorrelated point source,
    \begin{align}\label{eq:gammaphiu}
      \phi_u(r) &= \frac{X_u(r)}{4 \pi r^2} + \Finvthree\left\{ \frac{c \left(\left(z^2+1\right) \tan ^{-1}(z)+z\right)^2}{2 z^2 \left(z^2+1\right)
   \left(-c+z^2+1\right)} \right\} \\
      &= \frac{e^{-r} (2+r)}{8 \pi  r^2}+\int_0^{\infty } \frac{c \left(z+\left(1+z^2\right) \tan
   ^{-1}(z)\right)^2 \sin (r z)}{4 \pi ^2 r z \left(1+z^2\right) \left(1-c+z^2\right)} \,
   dz.
    \end{align}
    In Figure~\ref{fig-collisionflux} we show several examples of the cross validation of Eqs.~(\ref{eq:gammaphic}) and~(\ref{eq:gammaphiu}) with Monte Carlo estimation using a generalized collision estimator for the flux integrals where at each collision the flux score is $(1+s)/s$, except for the first collision following uncorrelated emission, which scores $(2+s)/(1+s)$.

    \subsubsection{Plane-Source Validation}
    Given the simpler track-length calculation with a finite slab versus a finite shell, we chose to test the track-length estimators in Section~\ref{sec:MC} about an isotropic plane source in infinite 3D media.  Applying the point-to-plane transform~\cite{case53}
    \begin{equation}
      \phi_{\text{pl}}(x) = 2 \pi \int_{|x|}^\infty \phipt(r) \, r \, dr,
    \end{equation}
    to the Fourier inversion for 3D space (Eq.~\ref{eq:sphericalFourier} with $d = 3$) we find that any given Fourier-space distribution $\bar{g}(z)$ has a primal space inversion about an isotropic plane source given by
    \begin{equation}\label{eq:FourierPtoP}
      g(x) = \frac{1}{\pi} \int_0^\infty \bar{g}(z) \cos( x z) dz = \frac{1}{\pi} \int_{-\infty}^\infty g(z) \e^{-i z x}  dz
    \end{equation}
    which is simply the Fourier inversion of the even function $\bar{g}(z)$.  It happens that in any dimension $d$ Eq.~\ref{eq:FourierPtoP} converts the spherical-geometry Green's functions derived in this section to the related Green's function for an isotropically-emitting hyper-plane source.

\begin{figure}
      \centering
      \subfigure[Correlated emission (Equation~\ref{eq:phicplane})]{\hspace*{-1.6cm} \includegraphics[width=1.2\linewidth]{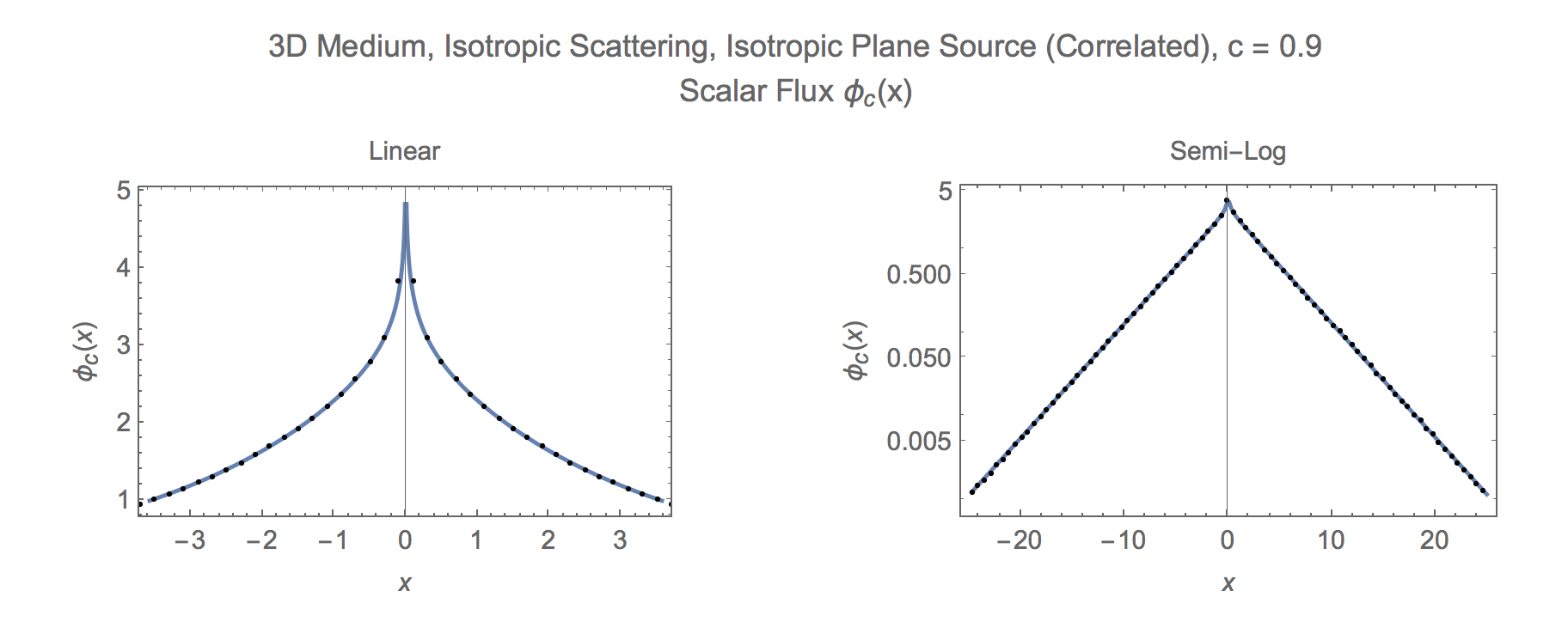}}
      \subfigure[Uncorrelated emission (Equation~\ref{eq:phiuplane})]{\hspace*{-1.6cm} \includegraphics[width=1.2\linewidth]{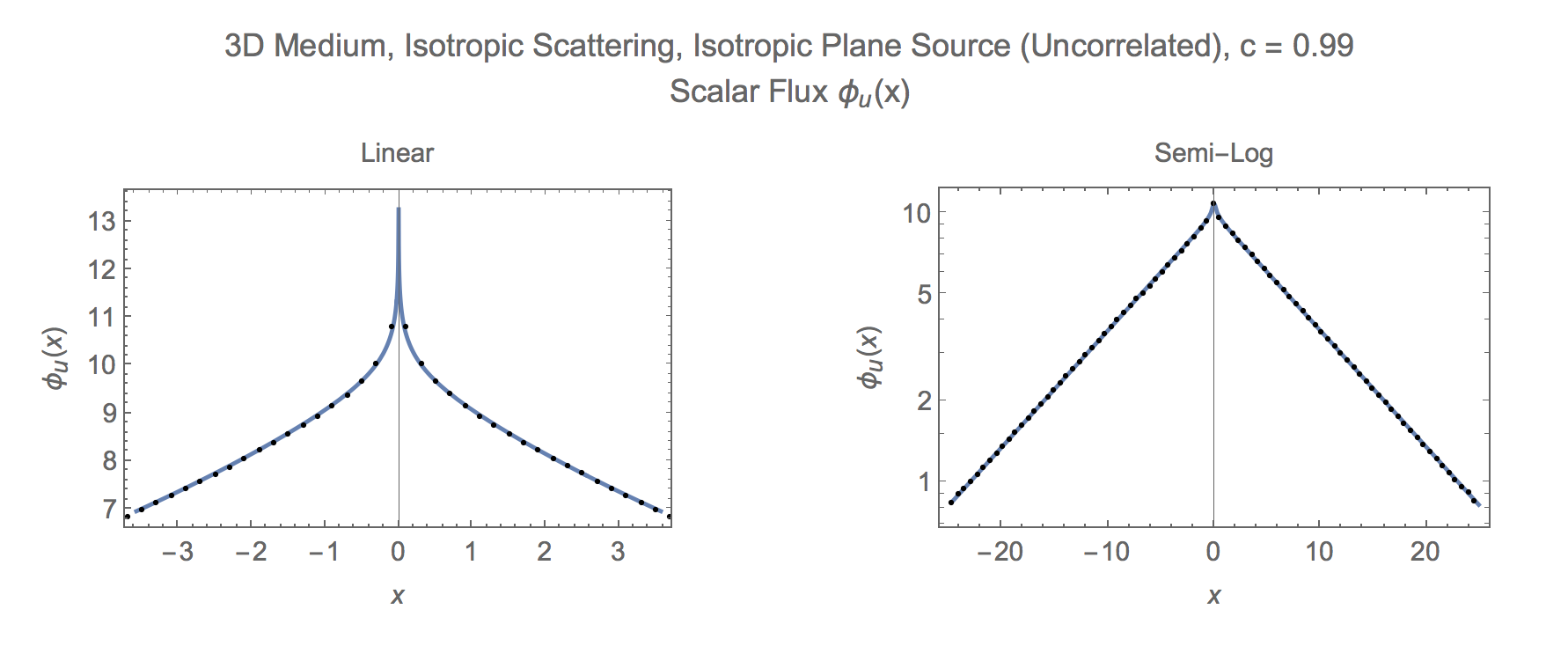}}
      \caption{Scalar flux density $\phi(z)$ at positions $z$ from an isotropic plane source in infinite 3D media.  Track-length Monte Carlo collision estimation (dots) vs analytic (continuous).}
      \label{fig-tracklengthflux} 
    \end{figure}

    Selecting $\bar{g}(z)$ to be the scalar flux for the correlated point source in Eq.~\ref{eq:gammaphic} and using Eq.~\ref{eq:FourierPtoP} we find the scalar flux about the isotropic plane source to be
    \begin{equation}\label{eq:phicplane}
      \phi_c(x) =  \frac{e^{-\left| x\right| }+E_1(\left| x\right| )}{2} 
          \, + \frac{1}{\pi } \int_0^{\infty } \frac{c \left(z+\left(1+z^2\right) \tan ^{-1}(z)\right) \cos (x z)}{ z \left(1+z^2\right) \left(1-c+z^2\right)} \, dz
    \end{equation}
    where $E_1$ is the exponential integral function $E_1(z) = \int_1^\infty \e^{-z t} / t dt$.  A similar calculation for the case of uncorrelated emission yields
    \begin{equation}\label{eq:phiuplane}
      \phi_u(x) = \frac{1}{4} \left(e^{-\left| x\right| }+2 E_1(\left| x\right|
   )\right)+  \frac{1}{\pi } \int_0^{\infty } \frac{\left(c \left(z+\left(1+z^2\right) \tan
   ^{-1}(z)\right)^2\right) \cos (x z)}{2 z^2 \left(1+z^2\right) \left(1-c+z^2\right)}
   \, dz.
    \end{equation}
    To validate these derivations, we performed track-length estimation of the flux integrals at various depths $z$ about the isotropic plane source, scoring track-lengths within each tally slab of thickness $dz = 0.2$.  Several comparisons of these deterministic and Monte Carlo results are shown in Figure~\ref{fig-tracklengthflux}.

\begin{figure}
      \centering
      \subfigure[Correlated emission (Equation~\ref{eq:Ccplane})]{\hspace*{-1.6cm} \includegraphics[width=1.2\linewidth]{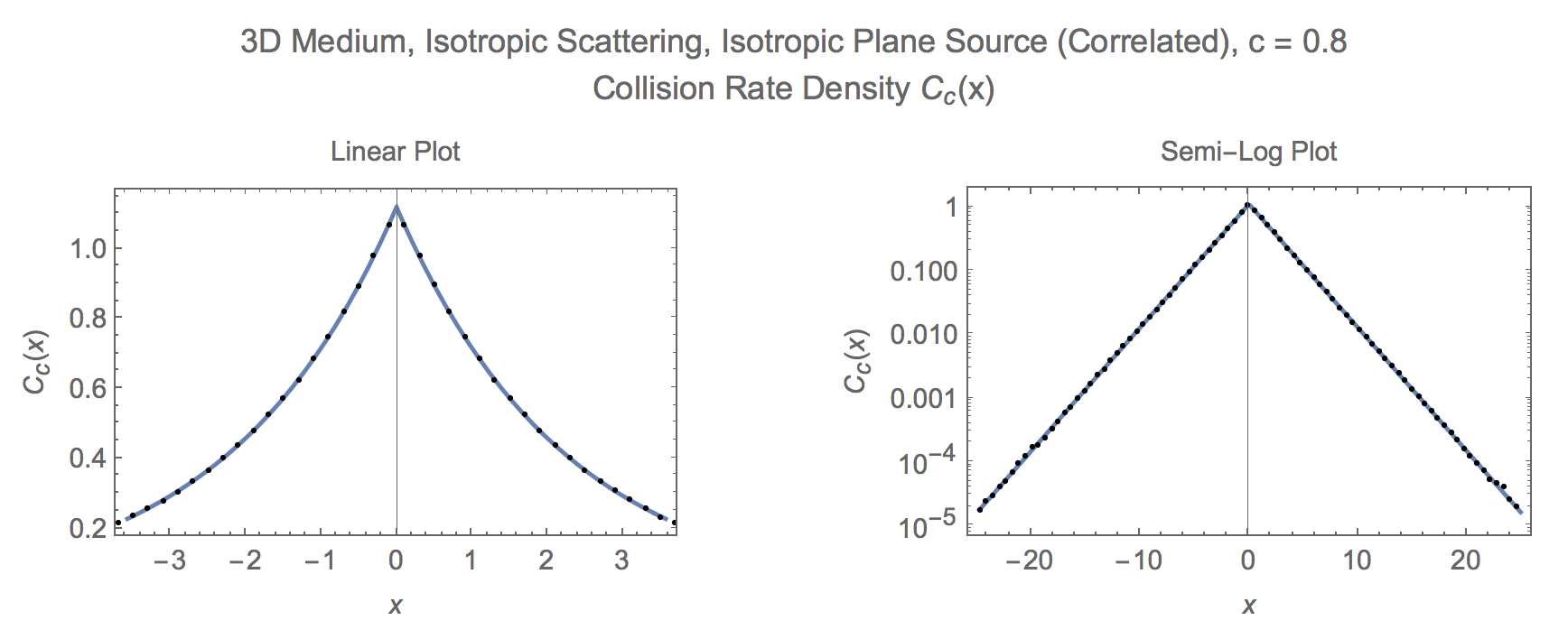}}
      \subfigure[Uncorrelated emission (Equation~\ref{eq:Cuplane})]{\hspace*{-1.6cm} \includegraphics[width=1.2\linewidth]{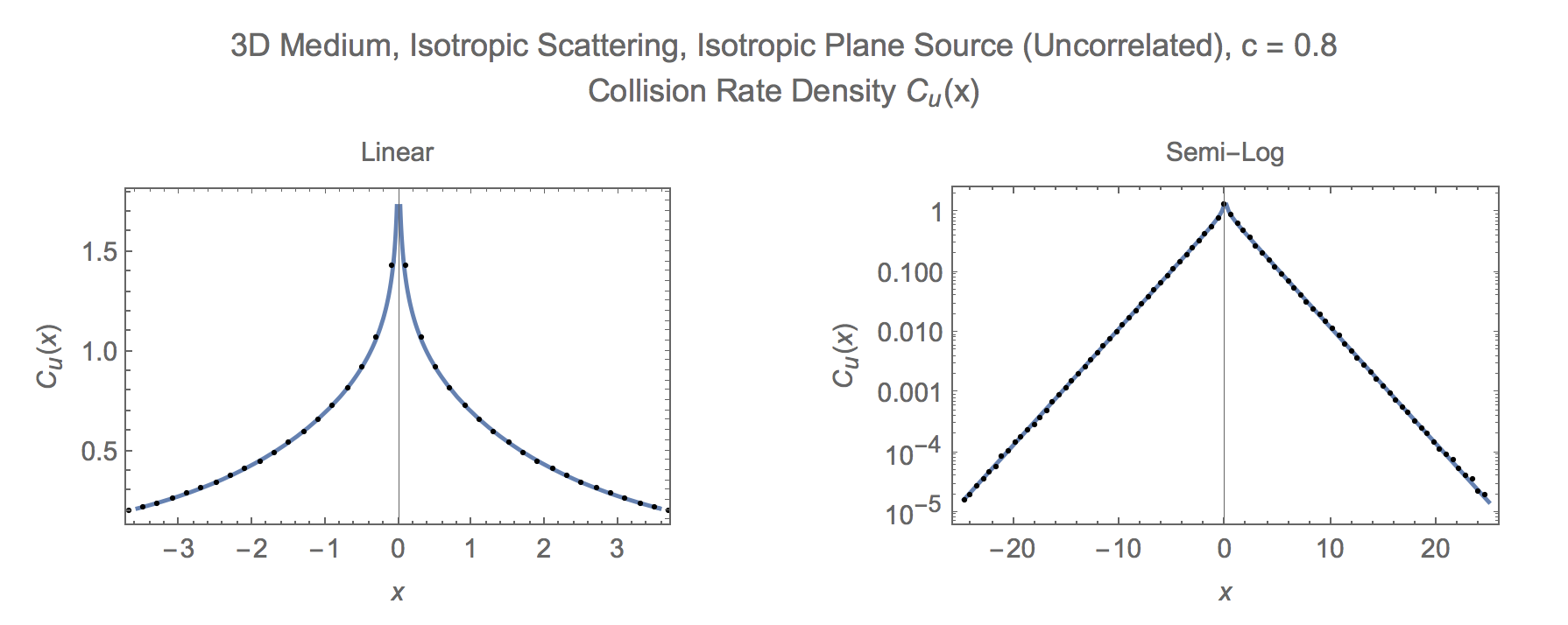}}
      \caption{Generalized track-length estimation of collision rate density at a distance $x$ from an isotropic plane source in infinite 3D media with isotropic scattering and correlated free-path distribution $p_c(s) = s \e^{-s}$.  Monte Carlo (dots) vs analytic (continuous).} 
      \label{fig-tracklengthcollision} 
    \end{figure}

    For completeness, we also estimated collision densities about the plane sources for both the correlated and uncorrelated emission cases using the new track-length estimators derived in Section~\ref{sec:tracklength} and validated the results using the Green's function derivations (Figure~\ref{fig-tracklengthcollision}).  As in the point-source case, we find a simple diffusion solution for the correlated emission case,
    \begin{equation}\label{eq:Ccplane}
      C_c(x) = \frac{1}{\pi} \int_0^\infty \frac{1}{1-c+z^2} \cos \left( x z \right)dz = \frac{e^{-\sqrt{1-c} \left| x\right| }}{2 \sqrt{1-c}},
    \end{equation}
    and the uncorrelated collision rate density is proportional to the correlated flux
    \begin{equation}\label{eq:Cuplane}
      C_u(x) = \frac{1}{2} \phi_c(x).
    \end{equation}
    For each free-path segment of the analog random walk, with $s_1$ and $s_2$ the range of intersection with a given flux tally volume element (satisfying $0 \leq s_1 \leq s_2 \leq s$), intercollision and correlated birth segments score the integrals previously given in Section~\ref{sec:TLintegrals}.

  \subsection{Flux / Collision-Rate Balance in Gamma-2 Infinite Media}

\begin{figure}
      \centering
      \subfigure[]{\includegraphics[width=.7\linewidth]{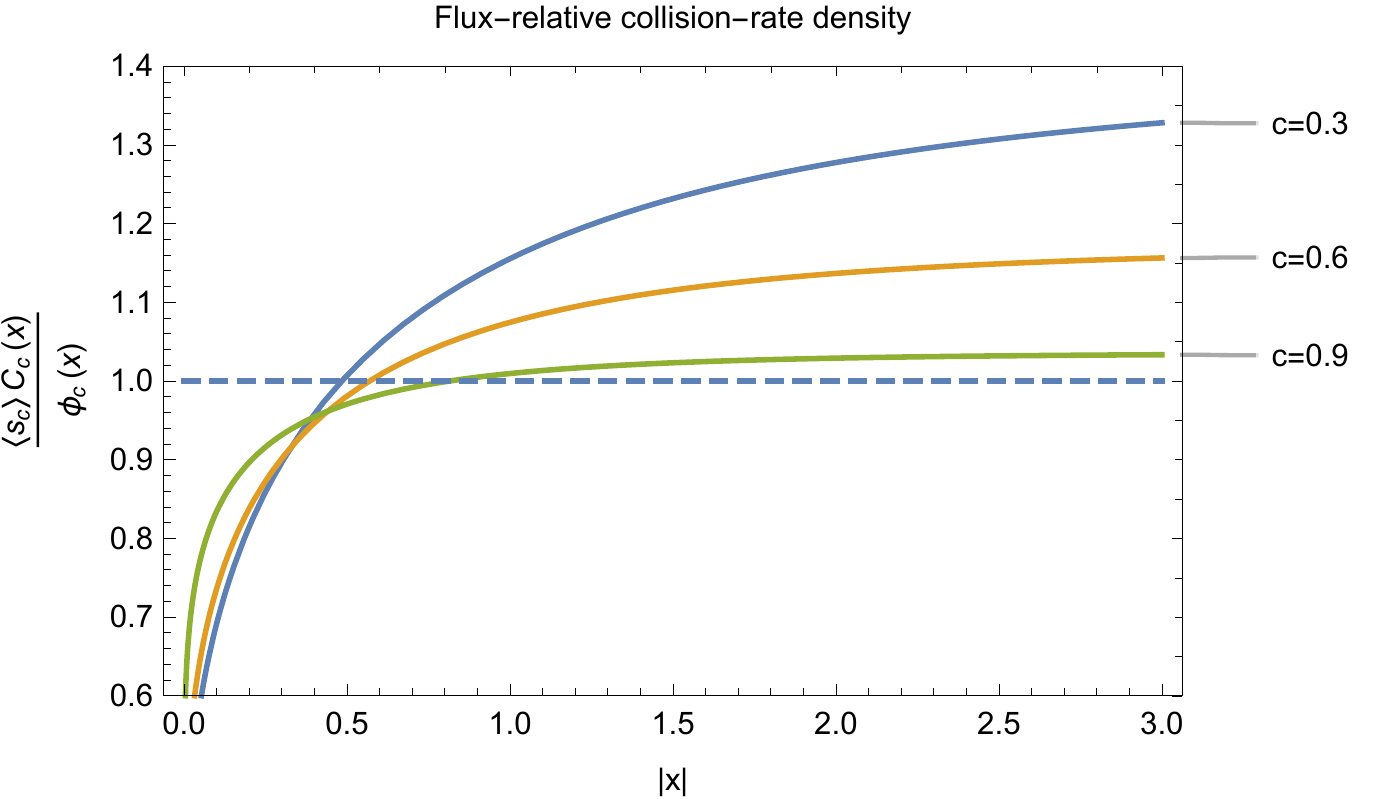}}
      \subfigure[]{\includegraphics[width=.7\linewidth]{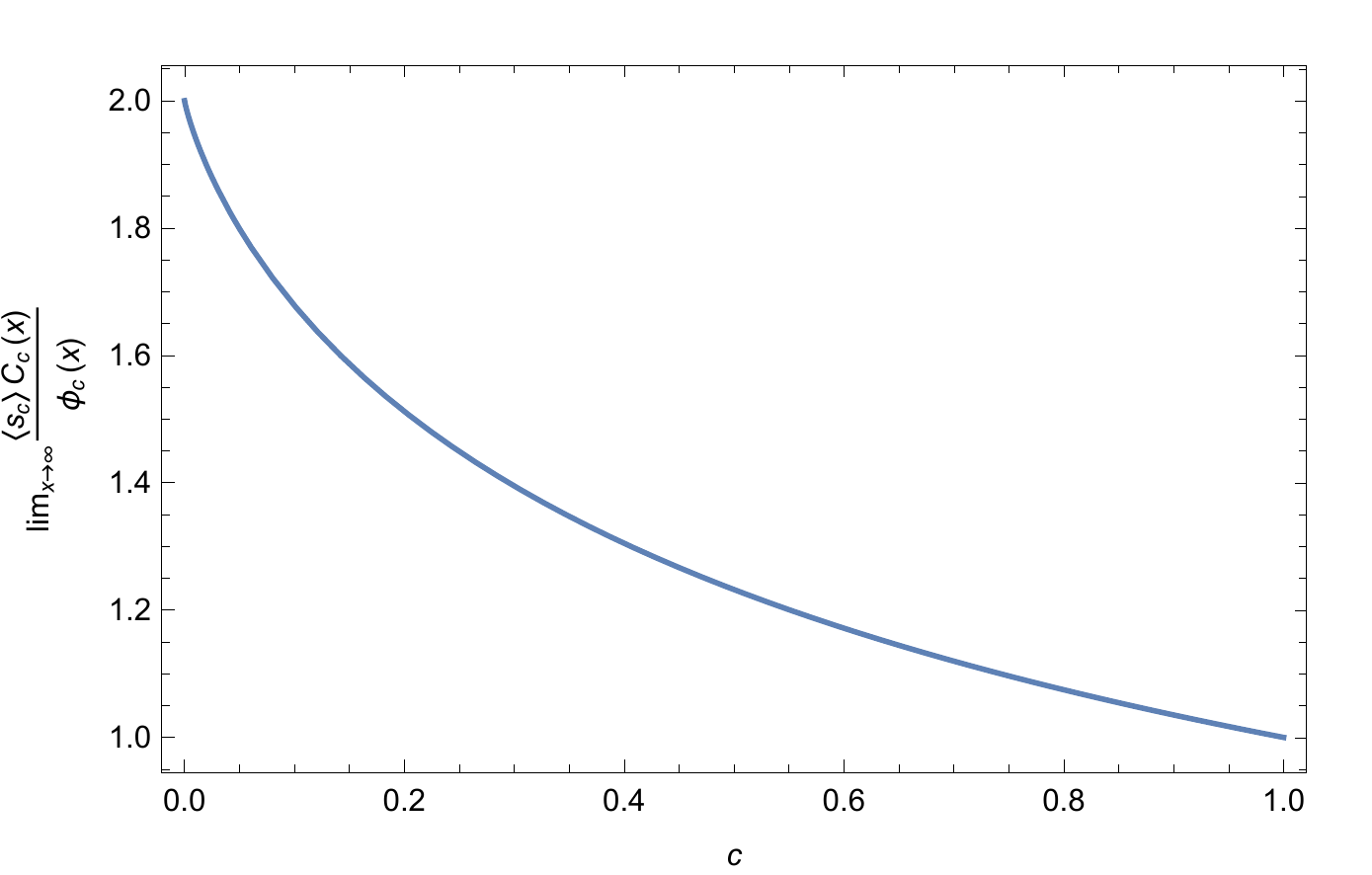}}
      \caption{\label{fig:IsoPlaneFluxCCRatioInf}For the correlated-emission isotropic plane source in infinite 3D media with correlated free-path distribution $p_c(s) = \e^{-s}s$ and single-scattering albedo $c$ we show (a) The flux-relative collision rate at distance $x$ from the source and (b) the limiting behaviour of the ratio at increasing distance from the source as a function of absorption (Equation~\ref{eq:limratio}).} 
    \end{figure}

    We briefly consider the ratio of collisions to particles in flight at various distances from the infinite isotropic plane source in the case of correlated emission, comparing the ratio to the classical proportionality  $\s C_c(x)  / \phi_c(x) = 1$.  Figure~\ref{fig:IsoPlaneFluxCCRatioInf} (a) illustrates that, for a variety of absorption levels, we see sub-classical collision rates near the plane source and super-classical collision rates as $x \rightarrow \infty$, with the classical balance occurring at a location that is increasingly far from the plane source as $c \rightarrow 1$.
    In the limit of increasing distance from the source we find that the ratio tends to a non-classical proportionality unless the scattering is conservative ($c = 1$),
    \begin{equation}\label{eq:limratio}
      \lim_{x \rightarrow \infty} \frac{\s C_c(x)}{\phi_c(x)} = \frac{2}{1+\frac{c \tanh ^{-1}\left(\sqrt{1-c}\right)}{\sqrt{1-c}}}.
    \end{equation}

    The ``rigorous'' asymptotic diffusion approximation for the correlated scalar flux is thus,
    \begin{equation}
      \phi_c(x) \approx \frac{\left(\sqrt{1-c}+c \tanh ^{-1}\left(\sqrt{1-c}\right)\right) }{2 (1-c)} e^{-\sqrt{1-c}
   \left| x\right| }
    \end{equation}

\section{Diffusion Approximations for GRT}\label{sec:diffusion}

\subsection{Free-path and transmittance moments}
Our diffusion approximations will require the first five spatial moments of the correlated-origin free path distribution $p_c(s)$, defined by
  	\begin{equation}
  		\langle s_c^m \rangle = \int_0^\infty p_c(s) s^m ds
  	\end{equation}
  	with $m \in \{ 1, 2, 3, 4 \}$.  The $m=0$ moment is unity by normalization of the distribution $p_c(s)$.  We will also require the 0th and 2nd spatial moments of $p_u(s), X_c(s)$ and $X_u(s)$, which can all be expressed in terms of the first five moments of $p_c(s)$ using the identity~\cite{vasques16}
  	\begin{equation}\label{eq:momentsmaster}
  		\langle s^m \rangle = m \int_0^\infty s^{m-1} X(s) ds
  	\end{equation}
  	where the uncorrelated moments are related to $s_c$ moments by $p_u(s) = X_c(s) / \s$.  Combining these, we find the moments for correlated transmittance,
  	\begin{align}\label{eqs:momentsXc}
  		&X_{c0} = \int_0^\infty X_c(s) ds = \s \\
  		&X_{c2} = \int_0^\infty X_c(s) s^2 ds = \frac{\sss}{3},
  	\end{align}
    the mean-square uncorrelated free-path
    \begin{equation}\label{eq:ssu}
      \ssu = \int_0^\infty p_u(s) s^2 ds = \frac{\sss}{3 \s}
    \end{equation}
    and the moments for uncorrelated transmittance,
    \begin{align}\label{eqs:momentsXu}
      &X_{u0} = \int_0^\infty X_u(s) ds = \frac{\ss}{2 \s} \\
      &X_{u2} = \int_0^\infty X_u(s) s^2 ds = \frac{\ssss}{12 \s}.
    \end{align}
    For a given scalar transport quantity $g(r)$ about an isotropic point source in infinite media, a moment-preserving diffusion approximation can be readily derived provided the moments
    \begin{equation}
       \langle r^0 g(r) \rangle = \int_0^\infty \Omega_d(r) r^0 g(r) dr
    \end{equation}
    and
    \begin{equation}
       \langle r^2 g(r) \rangle = \int_0^\infty \Omega_d(r) r^2 g(r) dr
    \end{equation}
    are known.  These moments are related to derivatives of the Fourier-transformed quantity $\bar{g}(z)$ at $z = 0$, as shown in Appendix A of~\cite{zoia11e}, by
    \begin{equation}\label{eq:zoia}
      \langle r^m g(r) \rangle = \frac{\sqrt{\pi}\Gamma \left( \frac{d+m}{2} \right)}{\Gamma \left( \frac{d}{2} \right) \Gamma \left( \frac{1+m}{2} \right)} \frac{\partial^m}{\partial(i z)^m} \left[ \bar{g}(z) \right]_{z=0}
    \end{equation}
    for even $m \geq 0$ and $\langle r^m g(r) \rangle = 0$ for odd $m$.  We only require this result for $m = 0$,
  	\begin{equation}\label{eq:zoia0}
  		\langle r^0 g(r) \rangle = \bar{g}(0)
  	\end{equation}
    and $m=2$,
  	\begin{equation}\label{eq:zoia2}
  		\langle r^2 g(r) \rangle = -d \bar{g}''(0).
  	\end{equation}

\subsection{Spherical Diffusion Mode in $d$ Dimensions}\label{sec:diffusionmodesphere}

    The classical steady-state diffusion equation, typically written
    \begin{equation}\label{eq:diffusion}
      - D \nabla^2 \phi(\pos) + \Sigma_a \phi(\pos) = Q(\pos)
    \end{equation}
    with diffusion constant $D > 0$ and constant $\Sigma_a > 0$ has simple solutions in infinite homogeneous media \newtwo{for the} isotropic point source $Q(\pos) = \delta(\pos)$.  The point-source diffusion mode in spatial dimension $d$ with diffusion length $\nu = \sqrt{D / \Sigma_a} > 0$ is the inverse spherical Fourier transform
      \begin{equation}
        m_d(\nu,r) = \Finv \left\{ \frac{1}{1+(z \nu)^2} \right\} = (2 \pi )^{-d/2} r^{1-\frac{d}{2}} \nu^{-\frac{d}{2}-1}
   K_{\frac{d-2}{2}}\left(\frac{r}{\nu}\right)
      \end{equation}
      where $K_n(x)$ is the modified Bessel function of the second kind.  The radial diffusion mode for the one-dimensional rod ($d = 1$) is
      \begin{equation}
        m_1(\nu,r) = \frac{e^{-\frac{r}{\nu}}}{2 \nu},
      \end{equation}
      for Flatland ($d = 2$) is
      \begin{equation}
        m_2(\nu,r) = \frac{K_0\left(\frac{r}{\nu}\right)}{2 \pi  \nu^2},
      \end{equation}
      and for three dimensions is
      \begin{equation}
        m_3(\nu,r) = \frac{e^{-\frac{r}{\nu}}}{4 \pi  r \nu^2}.
      \end{equation}

      We chose this form for its unit normalization,
      \begin{equation}\label{eq:mom1}
      	\int_0^\infty \Omega_d(r) m_d(\nu,r) = 1,
      \end{equation}
      yielding a second spatial moment of
      \begin{equation}\label{eq:mom2}
      	\int_0^\infty r^2 \, \Omega_d(r) m_d(\nu,r) = 2 d \nu^2.
      \end{equation}
      Our diffusion approximations of a given radially-symmetric scalar quantity $g(r)$ with moments $\langle r^0 g(r) \rangle$ and $\langle r^2 g(r) \rangle$ will be determined directly by approximating the quantity by a diffusion mode with two unknowns $a$ and $\nu$,
      \begin{equation}
        g(r) \approx a \, m_d(\nu,r)
      \end{equation} 
      and solving for $a$ and $\nu$ given the moments and Eqs.(\ref{eq:mom1}) and~(\ref{eq:mom2}), giving\begin{equation}\label{eq:diffusionfit}
        g(r) \approx \langle r^0 g(r) \rangle m_d\left(\sqrt{\frac{\langle r^2 g(r) \rangle}{2 d \langle r^0 g(r) \rangle }},r\right).
      \end{equation}
      Much of the complexity of the approximation is in finding the moments, which we solve generally in the remainder of this section.  These moment relations should also allow other forms of approximation using two-parameter families of radially-symmetric functions distinct from the diffusion mode.  Extending the following derivations to include higher-order moments is straightforward.  In the future we hope to investigate the utility of powers of diffusion in frequency space with power $p \ge 1/2$, which have simple inversions in all dimensions~\cite{jakeman1976model}
      \begin{equation}
        \Finv \left\{ \left(  \frac{1}{1+ (z\nu)^2} \right)^p \right\} = \frac{\pi ^{-d/2} 2^{-\frac{d}{2}-p+1} \nu ^{-\frac{d}{2}-p} r^{p-\frac{d}{2}}
   K_{\frac{1}{2} (d-2 p)}\left(\frac{r}{\nu }\right)}{\Gamma (p)}
      \end{equation}
      and permit more general fitting of exact transport solutions in either spherical or planar geometries.

  \subsection{Collision Rate Density}
	  
	  The zeroth moment of the collision rate density $\langle r^0 C(r) \rangle$ is a uniform integration of collision rate density over all space, which must then equal the mean number of collisions experienced by a single particle given an absorption probability of $1-c$ at each collision~\new{\cite{ivanov1994resolvent}}
      \begin{equation}\label{eq:C0}
        \langle r^0 C_c(r) \rangle = \langle r^0 C_u(r) \rangle =  \frac{1}{1-c},
      \end{equation}
	  a quantity that is independent of the phase function, free-path distribution(s), spatial dimension or angular distribution of emission.  Given its independence on free-path distribution, this quantity is the same for correlated and uncorrelated emission.

	  \subsubsection{Correlated emission}

	  	For the classical approach of approximating the entire collision rate density with a single diffusion mode, we require $\langle r^2 C_c(r) \rangle$ which, by Eqs.~\ref{eq:Cc} and \ref{eq:zoia2}, is
      \begin{equation}
        \langle r^2 C_c(r) \rangle = -d \, \frac{\partial^2}{\partial z^2} \left[ \frac{\bar{\zeta}_c(z)}{1 - c \bar{\zeta}_c(z)} \right]_{z=0} =  \frac{d \left((1-c \bar{\zeta}_c (0)) \bar{\zeta}_c ''(0)+2 c \bar{\zeta}_c '(0)^2\right)}{(c \bar{\zeta}_c (0)-1)^3}.
      \end{equation}
      From normalization of the propagator we know $\bar{\zeta}_c (0) = 1$.  Additionally, odd derivatives at the origin are $0$, $\bar{\zeta}_c' (0) = 0$.  Finally, from Eq.\ref{eq:zoia2} we have
      \begin{equation}
        \bar{\zeta}_c'' (0) = \frac{-\ss }{d}.
      \end{equation}
      Thus,
      \begin{equation}\label{eq:CCr2}
        \langle r^2 C_c(r) \rangle = \frac{\ss}{(1-c)^2}.
      \end{equation}
      With the two moments in Eqs.~(\ref{eq:C0}) and (\ref{eq:CCr2}), Eq.(\ref{eq:diffusionfit}) gives the full diffusion approximation
		  \begin{equation}
		  	C_c(r) \approx \frac{1}{1-c} m_d\left( \sqrt{\frac{\ss}{2 d (1-c)}} ,r \right).
		  \end{equation}
      This is then the Green's function for the diffusion equation for collision rate density under correlated emission
      \begin{equation}
        - D_{C_c} \nabla^2 C_c(\pos) + \frac{1}{1-c} C_c(\pos) = Q_c(\pos)
      \end{equation}
      with diffusion constant
      \begin{equation}\label{eq:DCc}
        D_{C_c} = \frac{\nu^2}{\langle r^0 C_c(r) \rangle} = \frac{\ss}{2 d}.
      \end{equation}
      It was previously conjectured~\cite{asadzadeh08} that the diffusion constant has a dimensional-dependence of $\frac{1}{d}$, which we previously \new{proved} for the case of exponential free paths and conjectured to hold for the case of general free-path distribution~\cite{deon14}.  We have proven this conjecture here and in the following derivations we find it also holds for uncorrelated emission and for moment-preserving diffusion approximations of the scalar flux as well.

  		We also generalize the approach of Grosjean~\shortcite{grosjean56a} of separating the density of first collisions from the multiply-scattered portion and only using the diffusion mode for approximating the latter.  The density of initial collisions has a simple analytic form and, further, it corresponds precisely to the portion of the total vector collision rate density (for particles entering collisions) that is singular in angle---in direct conflict with the angular smoothness of diffusion solutions.  For these reasons, Grosjean chose to separate transport quantities into two parts, approximating one with a diffusion equation and handling the initial collision rates and uncollided flux directly, noting improved accuracy in many cases for calculations involving high absorption levels and in locations near sources and boundaries.  Further details on his findings and application of the approach to convex bounded domains can be found in~\cite{grosjean56a,grosjean56b,grosjean58a,grosjean58b}.

      To form the diffusion approximation for the multiply-collided collision density
      \begin{equation}
        G(r) = C_c(r) - \frac{ p_c(r)}{\Omega_d(r)},
      \end{equation} we require the expected number of multiple collisions which, given that the first collision always happens with probability $1$, is
      \begin{equation}
        \langle r^0 G(r) \rangle = \frac{1}{1-c} - 1 = \frac{c}{1-c}.
      \end{equation}
      The second moment is found via Eq.(\ref{eq:Cc2}) and (\ref{eq:zoia2}),
      \begin{equation}
          \langle r^2 G(r) \rangle = -d \, \frac{\partial^2}{\partial z^2} \left[ \frac{c \, ( \bar{\zeta}_c(z))^2}{1 - c \bar{\zeta}_c(z)} \right]_{z=0} = \frac{(2-c) \, c \, \ss}{(1-c)^2},
      \end{equation}
      giving the complete approximation,
		  \begin{equation}
		  	C_c(r) \approx \frac{ p_c(r)}{\Omega_d(r)} +\frac{c}{1-c} m_d\left( \sqrt{\frac{(2-c) \ss}{2 d (1-c) }} ,r \right).
		  \end{equation}

      \subsubsection{Uncorrelated emission}

        The case of uncorrelated emission is treated in a similar fashion.  The normalization of the uncorrelated free-path distribution gives $\bar{\zeta}_u (0) = 1$.  Combining Eqs.(\ref{eq:ssu}) and (\ref{eq:zoia2}) gives
        \begin{equation}
          \bar{\zeta}_u'' (0) = \frac{-\sss}{3 \, \s \, d}.
        \end{equation}
        Using Eqs.(\ref{eq:Cu}) we find
        \begin{equation}\label{eq:CUr2}
          \langle r^2 C_u(r) \rangle = \frac{3 c \s \ss + \sss (1-c)}{3 (1-c)^2 \s}.
        \end{equation}
  	  	Thus, the classical diffusion approximation for uncorrelated emission is
  		  \begin{equation}
  		  	C_u(r) \approx \frac{1}{1-c} m_d\left( \sqrt{\frac{3 c \s \ss + \sss(1-c)}{6 d \s (1-c)}} ,r \right).
  		  \end{equation}	

        From Eq.(\ref{eq:Cu2}) we find
        \begin{equation}
          \langle r^2 G(r) \rangle = \frac{c\left(3  \s \ss + \sss (1-c)\right)}{3 (1-c)^2 \s}
        \end{equation}
        giving the Grosjean-form approximation,
  		  \begin{equation}
  		  	C_u(r) \approx \frac{ p_u(r)}{\Omega_d(r)} +\frac{c}{1-c} m_d\left( \sqrt{\frac{3 \s \ss + \sss(1-c)}{6 d \s (1-c)}} ,r \right).
  		  \end{equation}

  \subsection{Fluence / Scalar Flux}

    For the case of scalar flux we take a similar approach but preserve the first two even spatial moments of a different physical transport quantity, the scalar flux, which is not generally proportional to the collision rate density.

    \subsubsection{Correlated emission}

      In the case of correlated emission, the track-lengths of all segments in the random walk have the same expected value $\s$ and are all independent random variables.  The 0th moment of the scalar flux is simply the expected track-length of the entire walk over all space.  With survival probability $0 < c < 1$ at each collision, the expected number of path segments in the walk is $1/(1-c)$, each with a mean length of $\s$, thus the 0th moment of the scalar flux is simply
      \begin{equation}
        \langle r^0 \phi_c(r) \rangle = \frac{\s}{1-c}.
      \end{equation}
      While the scalar flux and collision rate density do not generally exhibit the classical proportionality, we see that in the case of correlated emission in infinite media the 0th moment of the two quantities does show a familiar form,
      \begin{equation}
        \frac{\langle r^0 C_c(r) \rangle}{\langle r^0 \phi_c(r) \rangle} = \frac{1}{\s}.
      \end{equation}
      The complete approximations for correlated flux are found most readily from Eq.~\ref{eq:phiCbalance}, with the same diffusion lengths as the case of uncorrelated collision density, and a scale factor of $\s$, producing the 
      classical-form diffusion approximation
      \begin{equation}
      	\phi_c(r) \approx \frac{\s}{1-c} m_d\left( \sqrt{\frac{3 c \s \ss + \sss(1-c)}{6 d \s (1-c)}} ,r \right).
      \end{equation}
      and the 
      Grosjean-form approximation
      \begin{equation}
      	\phi_c(r) \approx \frac{ X_c(r)}{\Omega_d(r)} + \frac{c \s}{1-c} m_d\left(\sqrt{\frac{3 \s \ss + \sss(1-c)}{6 d \s (1-c)}} ,r \right).
      \end{equation}

    \subsubsection{Uncorrelated emission}
      In the case of uncorrelated emission, the expected track-length of the entire path is 
      \begin{equation}
        \langle r^0 \phi_u(r) \rangle = \langle r^0 \phi_c(r) \rangle - \s + \langle s_u \rangle = \frac{c \s}{1-c} + \frac{\ss}{2 \s},
      \end{equation}
      which is the mean total track-length for correlated emission with the correlated free-path mean for the initial step removed and replaced with the uncorrelated mean for that emission step.  The second moment can be determined from Eqs.(\ref{eq:phiu}) and (\ref{eq:zoia2})
      \begin{multline}\label{eq:reduceme}
          \langle r^2 \phi_u(r) \rangle = -d \, \frac{\partial^2}{\partial z^2} \left[ \bar{\chi}_u(z) + \bar{\chi}_c(z) \, c \, \frac{\bar{\zeta}_u(z)}{1 - c \bar{\zeta}_c(z)} \right]_{z=0} = \\ -d \left(\frac{c^2 \bar{\chi}_c(0) \bar{\zeta}_c''(0)}{(1-c)^2}+\frac{c \left(\bar{\chi}_c(0)
   \bar{\zeta}_u''(0)+\bar{\chi}_c''(0)\right)}{1-c}+\bar{\chi}_u''(0)\right)
      \end{multline}
      where here we have used $\bar{\zeta}_c'(0) = \bar{\chi}_c'(0) = 0$ and $\bar{\zeta}_c(0) = \bar{\zeta}_u(0) = 1$ to simplify.  From Eqs.(\ref{eqs:momentsXu}) we also have
      \begin{align}
        &\bar{\chi}_c''(0) = \frac{-\sss}{3 d} \\
        &\bar{\chi}_u''(0) = \frac{-\ssss}{12 \s d}
      \end{align}
      From Eq.(\ref{eq:momentsmaster}) we have $\bar{\chi}_c(0) = \s$ and we can reduce Eq.(\ref{eq:reduceme}) to
      \begin{equation}\label{eq:phiUr2}
        \langle r^2 \phi_u(r) \rangle = \frac{1}{12} \left(\frac{4 c (3 c \s \ss-2 c \sss+2
  \sss)}{(1-c)^2}+\frac{\ssss}{\s}\right)
      \end{equation}
      giving the full diffusion approximation
      \begin{equation}
      	\phi_u(r) \approx \left( \frac{c \s}{1-c} + \frac{\ss}{2 \s} \right) m_d\left( \sqrt{\frac{\frac{4 c (3 c \s \ss-2 c \sss+2
   \sss)}{(1-c)^2}+\frac{\ssss}{\s}}{24 \, d \left(\frac{c
   \s}{1-c}+\frac{\ss}{2 \s}\right)}} ,r \right).
      \end{equation}

    Finally, a straightforward calculation similar to the previous gives the Grosjean-form approximation,
      \begin{equation}
      	\phi_u(r) \approx \frac{ X_u(r)}{\Omega_d(r)} + \frac{c \s}{1-c} m_d\left( \sqrt{\frac{3 c \s \ss + 2 \sss(1-c)}{6 d \s (1-c)}} ,r \right).
      \end{equation}

    \subsection{Anisotropic Scattering}\label{sec:Daniso}

      \new{For simplicity, we have so far considered only isotropic scattering.  The Fourier-transform approach used in this section can be extended for the case of anisotropic scattering (persistent random walks) using the methods outlined by Grosjean~\shortcite{grosjean51,grosjean53}, but is considerably more involved.  For the case of collision rate about a correlated source, however, the required moments are known and we briefly review them here for completeness and to relate to previous work.}

      \subsubsection{Classical Diffusion Approximation for Anisotropic Scattering}

      \new{
      As noted previously, the $0$th moment of the collision rate is invariant to phase function,}
      \begin{equation}
        \langle r^0 C_c(r) \rangle = \frac{1}{1 - c}.
      \end{equation}
      \new{For random flights in Flatland with a general, symmetric phase function having mean cosine $-1 \le g \le 1$ the second moment of the $n$th collision is~\cite{hall1977amoeboid,kareiva1983analyzing}}
      \begin{equation}
        \langle r^2 C_c(r|n) \rangle = n \ss + \frac{g}{1-g} 2 \s^2  \left(n-\frac{1-g^n}{1-g}\right),
      \end{equation}
      \new{from which we can sum over all orders and introduce absorption to find}
      \begin{equation}\label{eq:Cm2aniso}
        \langle r^2 C_c(r) \rangle = \sum_{n=1}^\infty c^{n-1} \langle r^2 C_c(r|n) \rangle = \frac{\ss}{(1-c)^2} \left( 1 + c g\frac{2  \s^2 }{\ss (1-c g)} \right).
      \end{equation}
      \new{Equation~\ref{eq:Cm2aniso} agrees with a previous derivation for the case of exponential random flights in 3D~\cite{grosjean63b}\footnote{Grosjean~\shortcite{grosjean63b} (p.26) derives moment expressions for exponential free flights in 3D with general phase function including $\langle r^6 C_c(r) \rangle$ and $\langle r^8 C_c(r) \rangle$, but we find the latter expression to not match MC.} and is consistent with 2D and 3D results in plane geometry~\cite{gandjbakhche1992scaling}.  We suspect that Equation~\ref{eq:Cm2aniso} holds generally in $d$D.}

      \new{Combining the two moments with Equation~\ref{eq:diffusionfit}, we find the diffusion coefficient for correlated-emission collision rate generalizing Equation~\ref{eq:DCc} to anisotropic scattering,}
      \begin{equation}\label{eq:DCcaniso}
        D_{C_c} = \frac{\nu^2}{\langle r^0 C_c(r) \rangle} = \frac{\ss}{2 d} \left( 1 + c g\frac{2  \s^2 }{\ss (1-c g)} \right).
      \end{equation}

      \new{The scaling of a classical anisotropically-scattering media to an isotropically-scattering one based on the reduced scattering coefficient $\Sigma_s' = \Sigma_s (1-g)$, preserves the diffusion coefficient $D_{C_c}$.  This is known not to hold generally with nonexponential free paths~\cite{gandjbakhche1992scaling}, but GRT offers a much richer set of potential similarity relations by varying the free-path distribution as a whole, and not just its mean, and is an interesting area for future work.}

      \subsubsection{Grosjean Diffusion Approximation for Anisotropic Scattering}

      \new{For completeness, we include the Grosjean-form approximation for anisotropic scattering.  The $0$th moment of the multiply-collided collision density remains
      \begin{equation}
        \langle r^0 G(r) \rangle = \frac{1}{1-c} - 1 = \frac{c}{1-c}.
      \end{equation}
      The second moment is found from the Neumann series of the classical derivation above, removing the 1st term,
      \begin{equation}
          \langle r^2 G(r) \rangle = \sum_{n=2}^\infty c^{n-1} \langle r^2 C_c(r|n) \rangle = \frac{\ss}{(1-c)^2} \left( 1 + c g\frac{2  \s^2 }{\ss (1-c g)} \right)-\ss
      \end{equation}
      giving the complete approximation,
      \begin{equation}
        C_c(r) \approx \frac{ p_c(r)}{\Omega_d(r)} +\frac{c}{1-c} m_d\left( \sqrt{\frac{(2-c) \ss (1-c g)+2 g \s^2}{2 (1-c) d (1-c g)}} ,r \right).
      \end{equation}}

      \subsubsection{Related Parameter-Of-Smallness Diffusion Derivations}

      \new{Using a parameter-of-smallness asymptotic analysis, Larsen and Vasques~\shortcite{larsen11} derived a diffusion coefficient in 3D for the scalar flux for correlated emission that is related to Equation~\ref{eq:DCcaniso} by classical proportionality, $D_{\phi_c} = D_{C_c} / \s$, and so is not a moment-preserving approximation for the flux in general.  Frank and Goudon~\shortcite{frank10} derived a similar diffusion coefficient for general dimension $d$ that is the same as Larsen's when $c = 1$.  Exactly why these derivations preserve collision rate moments and not flux moments is not clear to us, but we suspect it relates to the transformed quantity in (6.4) of Larsen~\shortcite{larsen2007} that scales flux by the equilibrium distribution of free paths. This scaling includes the inverse mean free path $1 / \s$ which, in the classical case, converts the flux to a collision rate and also has a $O(\epsilon^{-1})$ scaling.  Frank and Goudon's result was also found by Rukolaine~\shortcite{rukolaine2016generalized}.  From a moment-preserving perspective (but for collision-rate, not flux), the absorption-dependent diffusion coefficient would seem to be preferred.}

      \new{To illustrate the potential severity of the mismatch between collision rate and flux asymptotics, consider correlated emission with the free-path distribution,
      \begin{equation}
        p_c(s) = \frac{12 s}{ (1+s)^5}.
      \end{equation}
      With a finite mean square free path,
      \begin{equation}
        \ss = \int_0^\infty p_c(r) r^2 dr = 3,
      \end{equation}
      the moment-preserving diffusion approximations follow and are well behaved.  For the scalar flux, however, given
      \begin{equation}
        X_c(s) = 1 - \int_0^s p_c(s') ds' = \frac{4 s+1}{(s+1)^4},
      \end{equation}
      we find, even for the uncollided flux alone, assuming no scattering,
      \begin{equation}
        \langle r^2 \phi(r) \rangle_{c=0} = \int_0^\infty X_c(r) r^2 dr = \infty
      \end{equation}
      and so a diffusion mode cannot preserve the 2nd moment of the scalar flux.  Given this difference, the flux diffusion asymptotics just mentioned~\cite{larsen11,frank10} are surprising.}

  \subsection{Discussion and Validation}

    \subsubsection{Reduction to the Classical Case}

      Compared to the simplicity of the classical $P_1$ diffusion approximation, the isotropic point source diffusion Green's function for the infinite medium has a much more complicated form in the generalized case, requiring moments up to $\ssss$.  In the case of uncorrelated emission, the scalar flux diffusion coefficient is
      \begin{equation}\label{eq:Dphiu}
        D_{\phi_u} = \frac{\nu^2}{\langle r^0 \phi_u(r) \rangle} =  \frac{\s \left(4 c \s (3 c \s \ss-2 c \sss+2
   \sss)+(1-c)^2 \ssss\right)}{6 d \left(2 c \s^2-c
   \ss+\ss\right)^2}.
      \end{equation}
      This result is, however, consistent with the classical diffusion ($P_1$) approximation.  In the case of classical scattering with correlated free-path distribution $p_c(s) = \Sigma_t \e^{-\Sigma_t s}$, the correlated moments are
      \begin{equation}
        \langle s_c^m \rangle = \frac{m!}{(\Sigma_t)^m}
      \end{equation}
      and the diffusion coefficient (\ref{eq:Dphiu}) reduces to the well-known
      \begin{equation}
        D_{\phi_c} = D_{\phi_u} = \frac{(1-c)^2}{\Sigma_t \, d} + \frac{(2-c)c}{\Sigma_t \, d} = \frac{1}{\Sigma_t \, d}
      \end{equation}
      with the portion $(1-c)^2 / (\Sigma_t \, d)$ arising from the uncollided flux and $(2-c)c / (\Sigma_t \, d)$ from the collided portion.

      \new{While there are effectively four forms of moment-preserving diffusion asymptotics in GRT, with two having identical decay rates, there is at most one asymptotic decay rate of the ``rigorous diffusion'' nature, corresponding to the largest \newtwo{real} eigenvalue (if one exists) of the collision kernel, a zero of the dispersion function~\cite{deon14}
      \begin{equation}
        \Lambda(z) = 1 - c \bar{\zeta}_c(i / z),
      \end{equation}
      which is a common denominator in all four frequency-space Green's function representations.
      } 

    \subsubsection{Power-Law Attenuation}

      We now consider the case of transport with a power-law attenuation law as defined earlier in Section~\ref{sec:TLintegrals}.  The full set of moments required for the diffusion approximations are
      \begin{align}
        &\s = \ell, \\
        &\ss = \ell^2 \frac{2 a }{a-1}, \\
        &\sss = \ell^3 \frac{6 a^2 }{a^2-3 a+2}, \\
        &\ssss = \ell^4 \frac{24 a^4 (a+1)  \Gamma (a-3)}{\Gamma (a+2)}
      \end{align}
      with the restrictions $a > m - 1$ for moment $\langle s_c^m \rangle$.

\begin{figure}
        \centering
        \subfigure[Correlated emission - $\langle r^2 C_c(r)\rangle$]{\includegraphics[width=.88\linewidth]{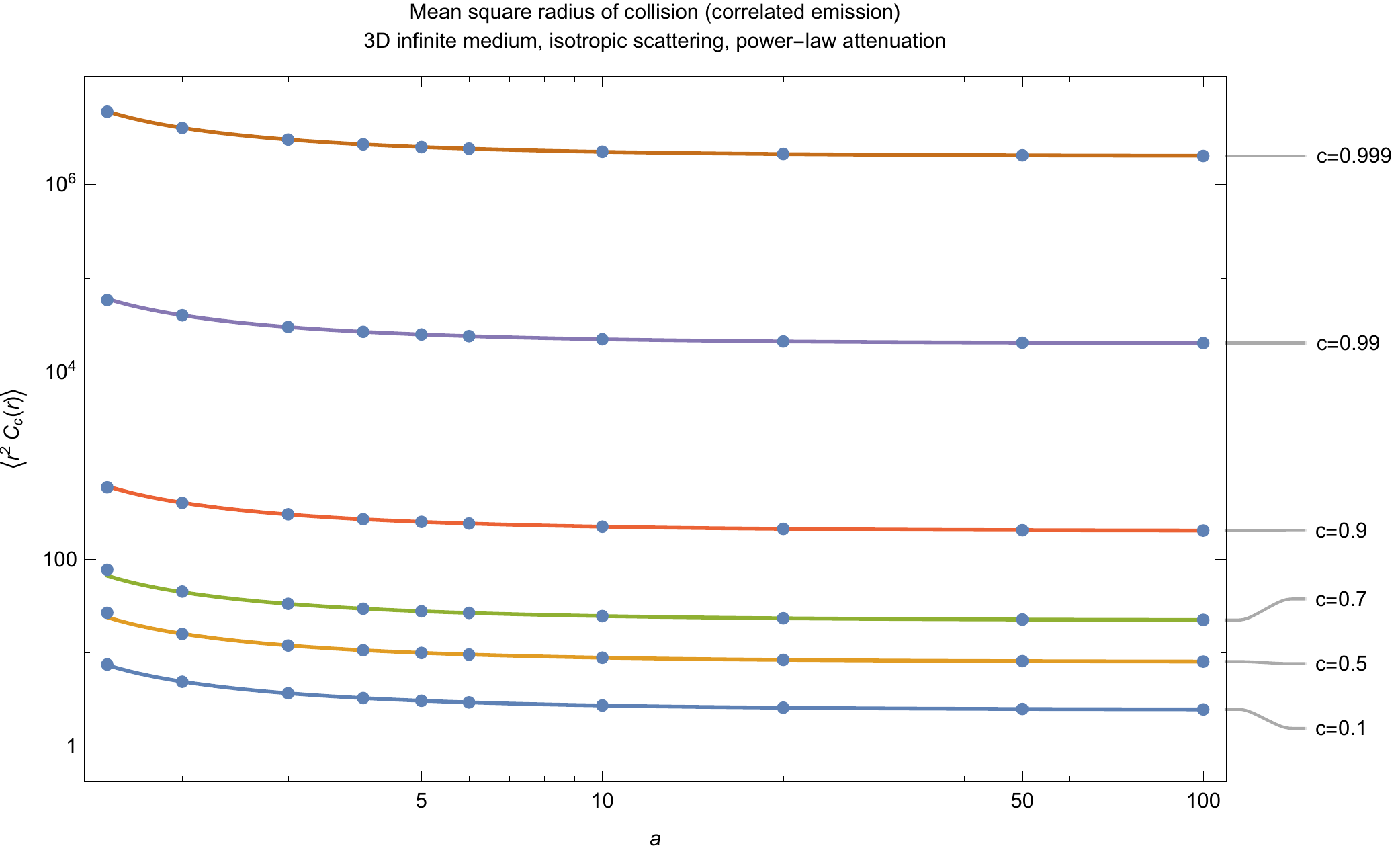}}
        \subfigure[Uncorrelated emission - $\langle r^2 C_u(r)\rangle$]{\includegraphics[width=.88\linewidth]{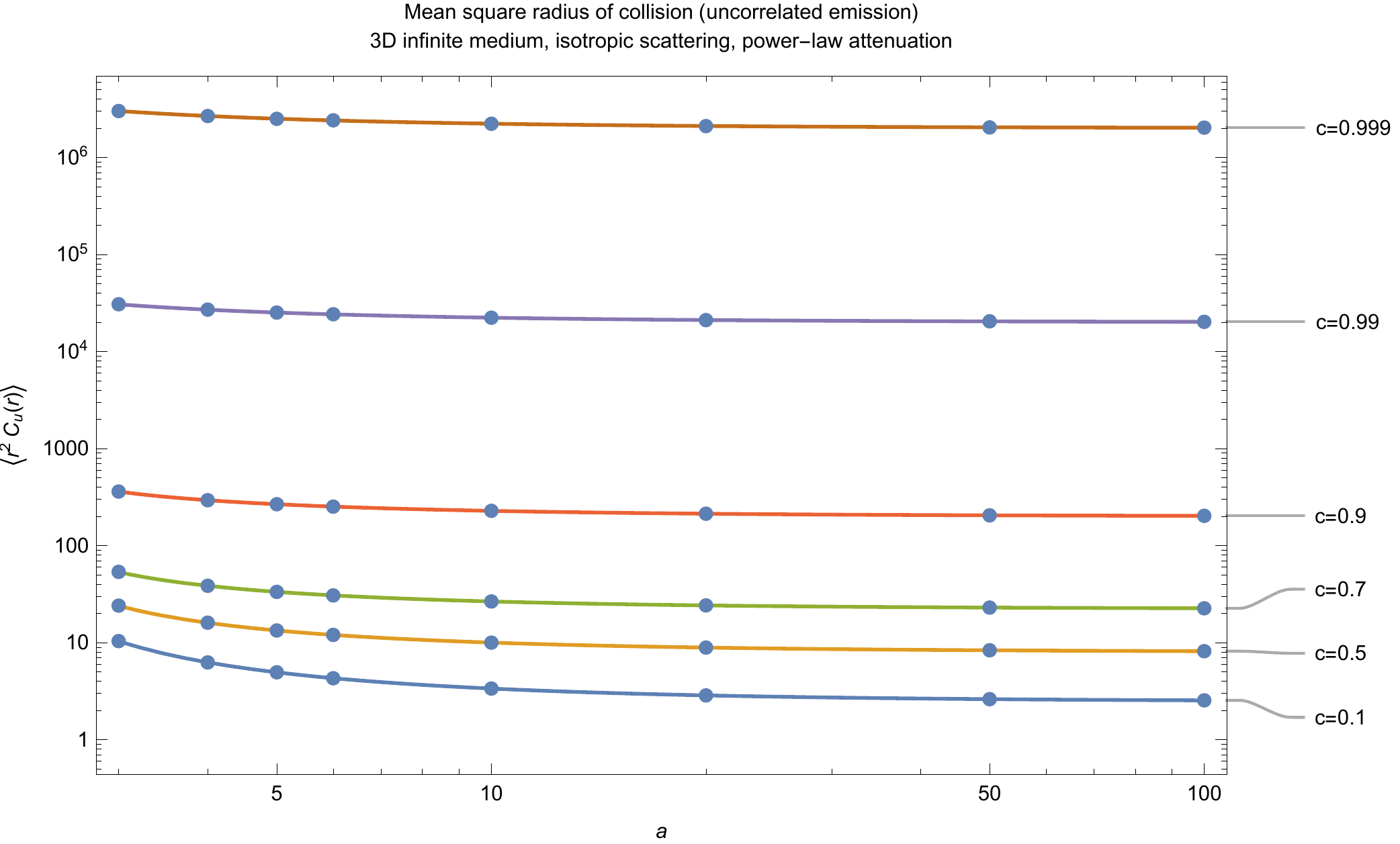}}
        \caption{\label{fig-C2moments}Mean square distance of collision in 3D for power-law transport: Monte Carlo (dots) vs. deterministic (continuous).} 
      \end{figure}

\begin{figure}
        \centering
        \subfigure[Correlated emission - $\langle r^2 \phi_c(r)\rangle$]{\includegraphics[width=.88\linewidth]{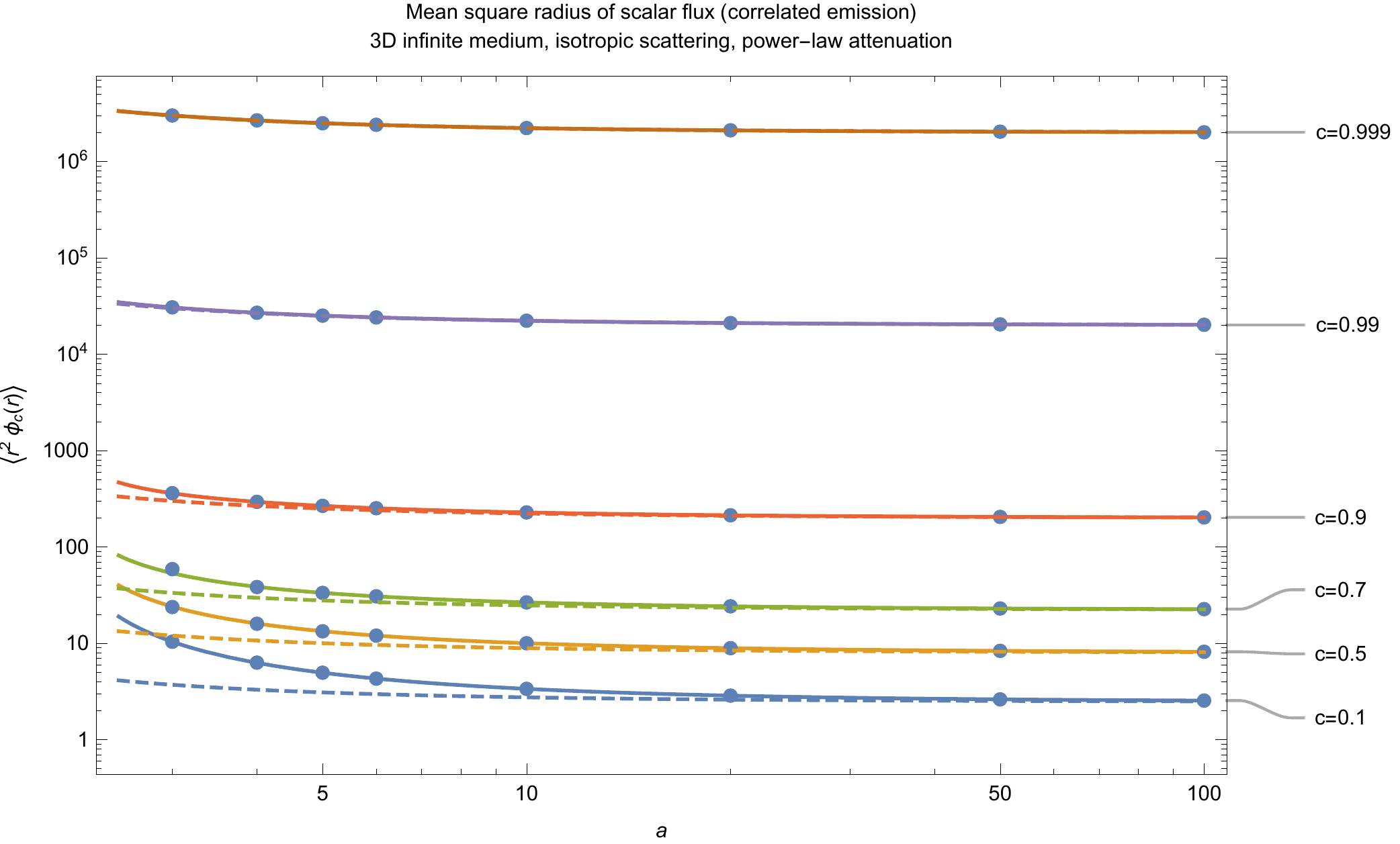}}
        \subfigure[Uncorrelated emission - $\langle r^2 \phi_u(r)\rangle$]{\includegraphics[width=.88\linewidth]{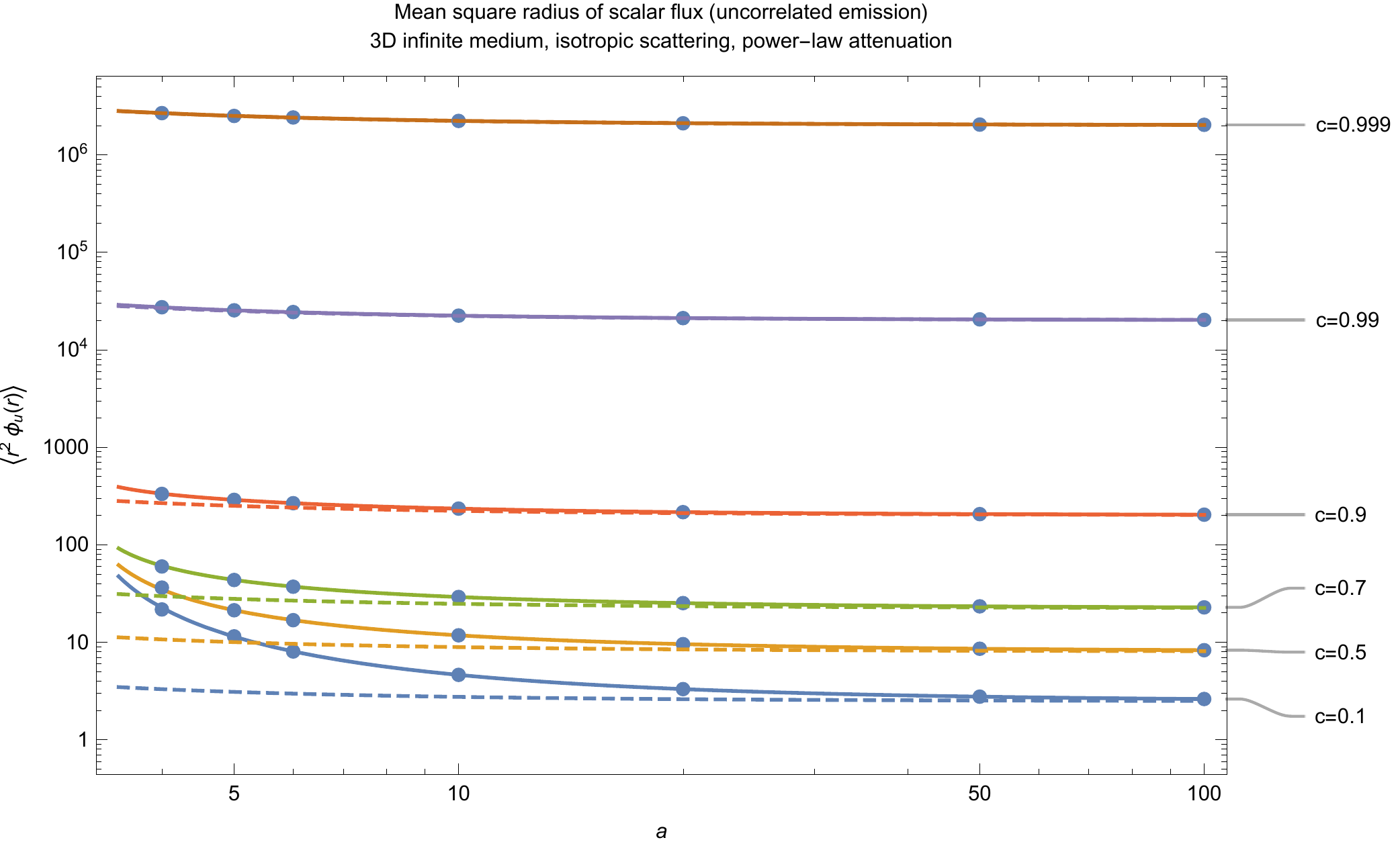}}
        \caption{\label{fig-phi2moments}Mean square scalar flux in 3D for power-law transport: Monte Carlo (dots) vs. deterministic (continuous).  The moments for collision rate density are shown (dashed) for comparison.} 
      \end{figure}

      We performed a suite of Monte Carlo simulations for power-law transport with unit mean correlated free path ($\s = \ell = 1$) about an isotropic point source in infinite 3D medium configurations.  For both the collision rate density and scalar flux we estimated tallies of both the radial quantities as well as their 2nd radial moments, all using an analog collision estimator.  For the moments, we compared Monte Carlo and deterministic predictions and found agreement in all cases.  For the collision density case, the Monte Carlo validation of Eqs.(\ref{eq:CCr2}) and (\ref{eq:CUr2}) is summarized in Figure~\ref{fig-C2moments}.  Likewise, Figure~\ref{fig-phi2moments} summarizes the validation of Eqs.(\ref{eq:CUr2}) and (\ref{eq:phiUr2}) for the scalar flux moments.  

      A comprehensive evaluation of the accuracy of the generalized diffusion approximations over a wide variety of free-path distributions is beyond the scope of the current work.  We briefly highlight some of our findings for the case of power-law transport and note several trends.  In a future part of this work we hope to expand upon this accuracy evaluation in the context of method-of-images application of these approximations to solve the searchlight problem in the style of~\cite{deon11a}.

\begin{figure}
        \centering
        \subfigure[$c = 0.5, a = 3$]{\includegraphics[width=.99\linewidth]{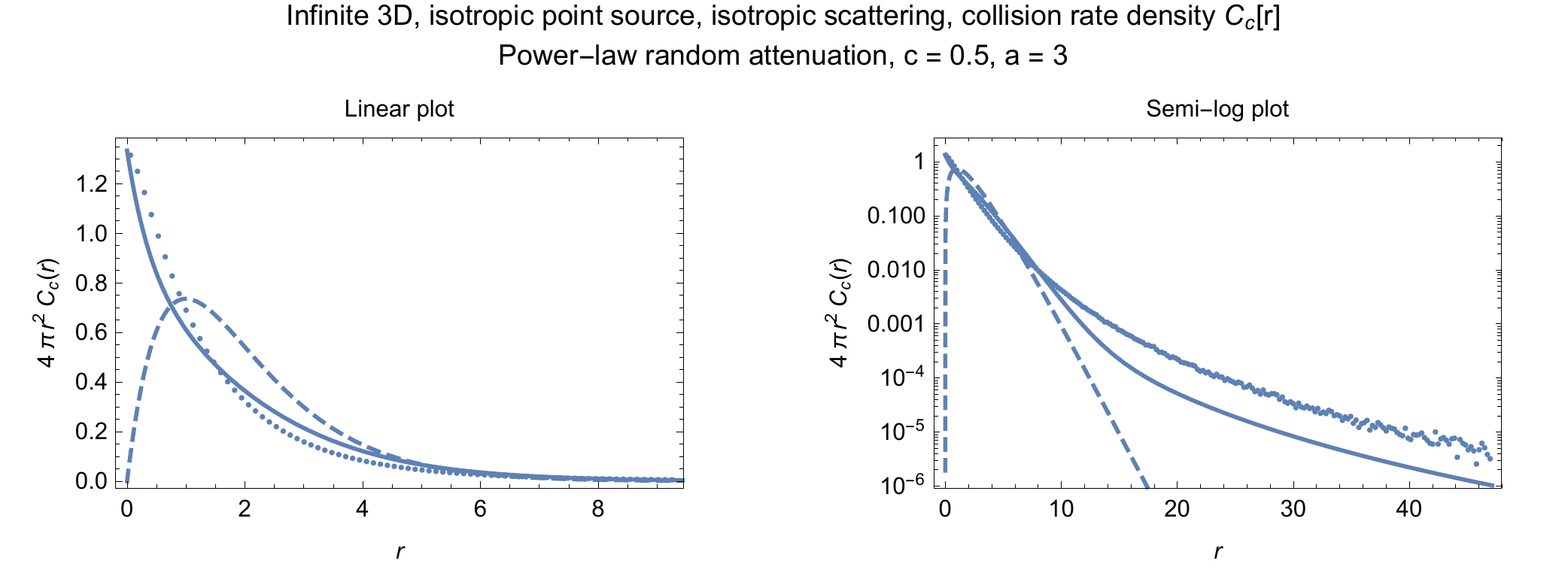}}
        \subfigure[$c = 0.9, a = 6$]{\includegraphics[width=.99\linewidth]{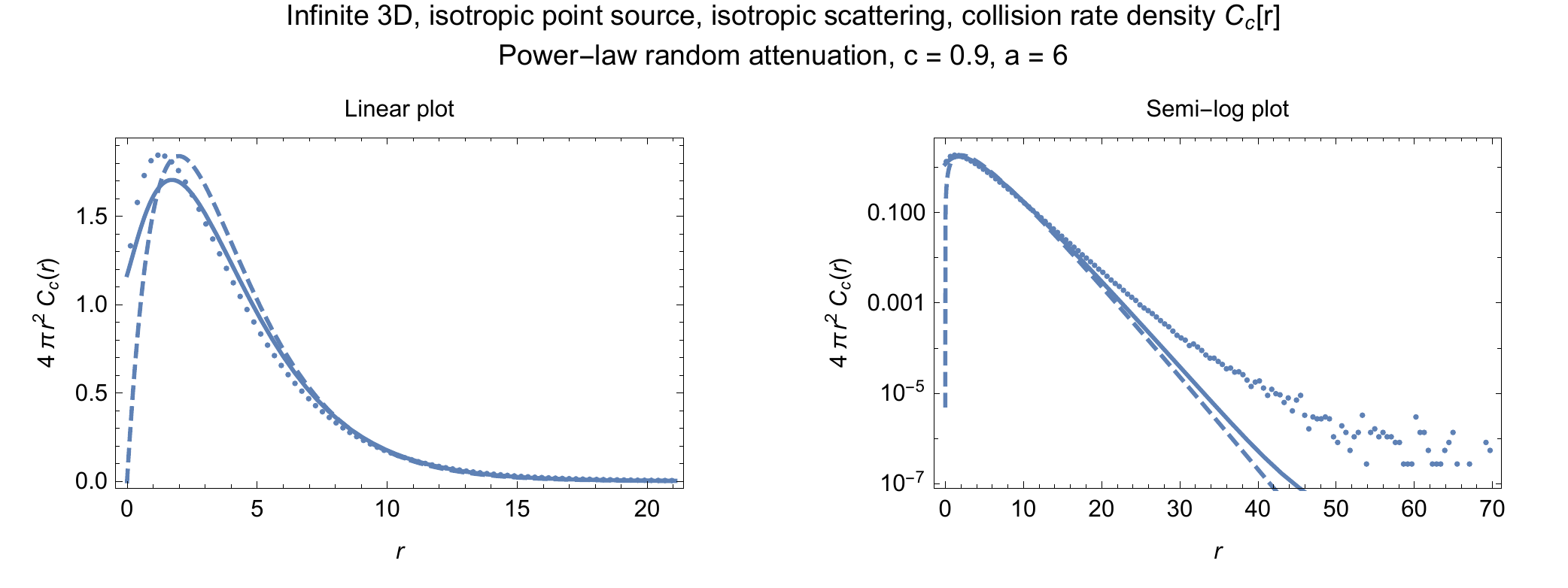}}
        \caption{\label{fig-Ccdiffusion}Collision-rate density $C_c(r)$ about an correlated-emission isotropic point as predicted by Monte Carlo (dots), diffusion approximation (dashed) and Grosjean-form diffusion approximation (continuous).} 
      \end{figure}

\begin{figure}
        \centering
        \subfigure[Collision-rate density $C_u(r)$]{\includegraphics[width=.99\linewidth]{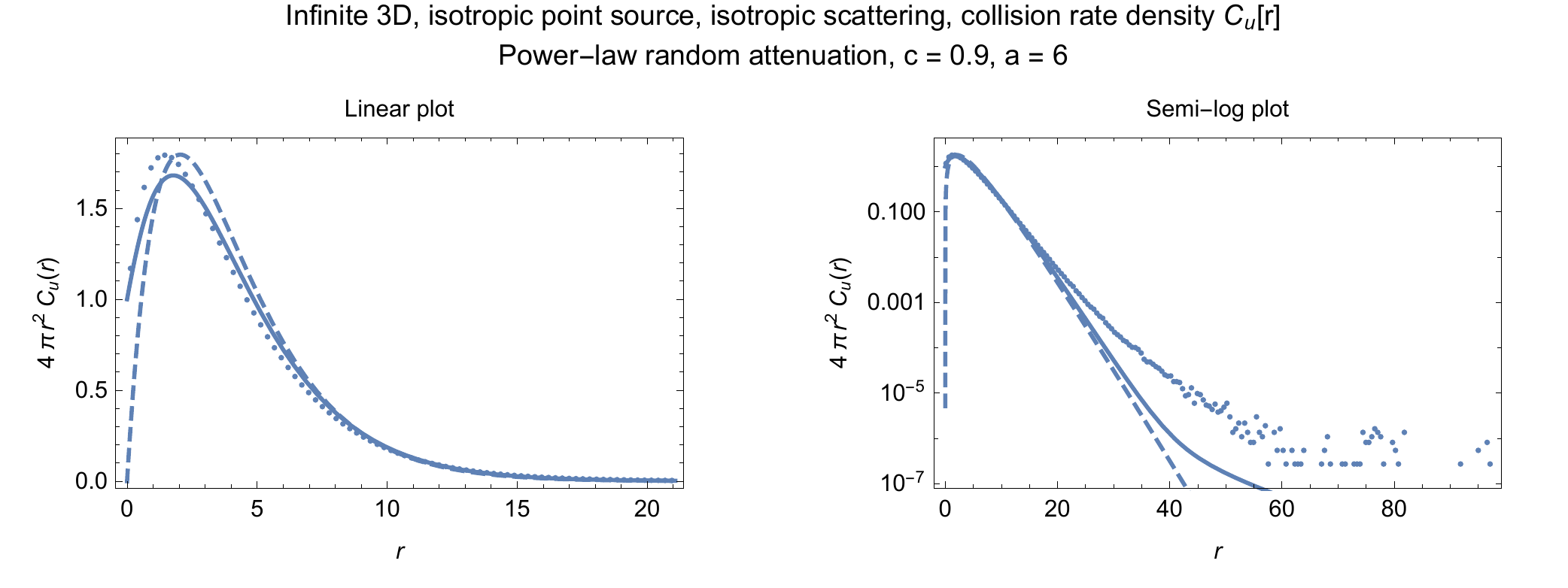}}
        \subfigure[Scalar Flux $\phi_u(r)$]{\includegraphics[width=.99\linewidth]{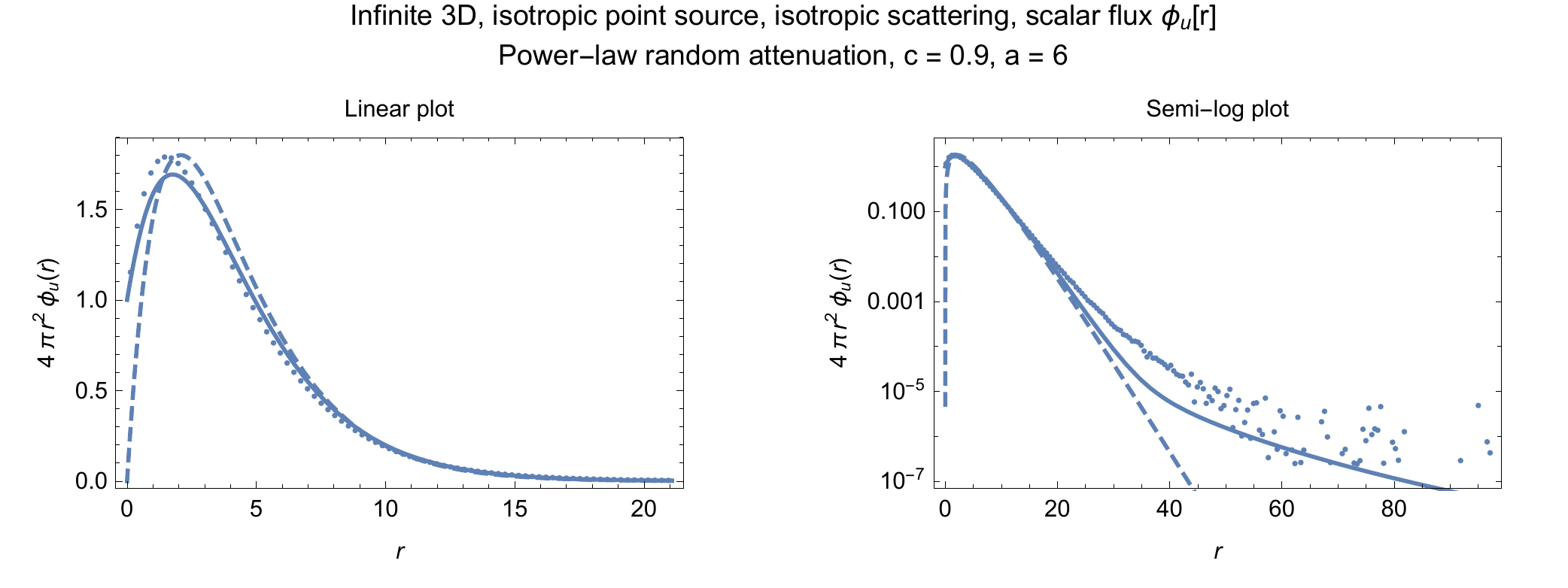}}
        \caption{\label{fig-Cuphiu} (a) Collision-rate density $C_u(r)$ and (b) scalar flux $\phi_u(r)$ for the case of uncorrelated emission from an isotropic point source in an infinite 3D medium with power-law attenuation with unit correlated mean free path $\s = \ell = 1$.  Predictions by Monte Carlo (dots), diffusion approximation (dashed) and Grosjean-form diffusion approximation (continuous).}
      \end{figure}

      Figures~\ref{fig-Ccdiffusion} and~\ref{fig-Cuphiu} show the performance of the diffusion and Grosjean-form diffusion approximations for collision rate density ($C_c(r)$ and $C_u(r)$), and scalar flux $\phi_u(r)$.  By Eq.(\ref{eq:phiCbalance}) the evaluation of $\phi_c(r)$ is equivalent to that of $C_u(r)$.  Some general trends from the case of classical exponential scattering survived the generalization to power-law transport; in the short range regime $r < 10 \s$, we noted lower overall relative error for the Grosjean-form approximation vs. the classical diffusion form in all cases.  By including the first term in the Neumann series solution exactly, the Grosjean-form approaches zero error everywhere as $c \rightarrow 0$, whereas the classical diffusion form remains consistently poor for very low radius ($r < \s$) regardless of absorption level.  For absorption levels in the range $0.5 < c < 0.9$ we saw five times less relative error for $r < 10 \s$ in the Grosjean case vs the classical case for near-classical transport ($a > 50$), with the Grosjean advantage shrinking to only a $20\%$ reduction in relative error when $a = 4$.  

      For absorption levels $c \leq 0.99$ we also noted a raised tail that is missing from the classical diffusion approximation and not a property of classical \new{transport}.  We attribute this to the free path distribution approaching a heavy-tailed distribution as $a \rightarrow 0$, with fewer bounded moments as $a$ decreases, thereby presenting an increasing challenge to the diffusion approximation as $a$ decreases.  This raised tail occurs at all absorption levels $c < 1$.  The uncollided term alone, having a power-law falloff, will eventually dominate any exponential, but it was not observable in an analog simulation at $c = 0.999$ with only $10^6$ \new{histories}.  The Grosjean-form approximation includes this bent tail, by including the power-law uncollided term exactly, but the relative error at these distances is seen to be as much as an order of magnitude.

\section{Discussion and Outlook}\label{sec:discussion}

  Under a theme of exploring the non-local proportionality between the density of particles in flight to the rate at which that flight leads to collisions, we have proposed new generalized Monte Carlo estimators, Green's functions and diffusion approximations for generalized linear Boltzmann transport.  An equilibrium imbedding of Larsen and Vasques's~\shortcite{larsen11} framework into bounded domains has required pairs of solutions in most cases, which were previously equivalent in the classical theory, and special handling of the two classes of free-paths throughout.  We distinguished between flux-proportional and collision-rate-proportional estimations and defined the analog walk for GRT.

  We found Larsen's formulation of GRT via $\Sigma_t(s)$ very powerful, making it straightforward to bring much of the machinery of classical Monte Carlo methods to bear.  Adjoint estimation, importance, and the complexity of bidirectional estimators~\cite{veach97,krycki2015mathematical,vitali2018comparison} and linear combinations of many forms of different estimators~\cite{krivanek2014unifying} has been left for future work.  

  We have preferred the integral form of the transport equation in GRT, wherever possible, to avoid handling $\psi(\pos,\dir,s)$ directly.  There are, however, some important cases we have not discussed where it would seem $\psi(\pos,\dir,s)$ is unavoidable.  To extend the condensed-history acceleration scheme of Fleck and Canfield~\shortcite{fleck84}, which uses precomputed or analytic scattering solutions in finite spherical domains, requires a notion of a virtual boundary in GRT---a boundary across which correlation is maintained.  In jumping to the surface of the virtual sphere boundary and carrying on the transport from there, knowledge of $\psi(\pos,\dir,s)$ at the boundary is required to ensure that the balance of collisions outside of the sphere is not changed by the acceleration.  This notion has been lacking from previous applications of the method to GRT~\cite{moon07,meng2015multi,muller2016efficient}, where $s$ has been reset to $0$ and correlated-origin free-path statistics follow each application of the accelerators.  \removetwo{In the study of invariance properties of random walks under certain assumptions in GRT, Mazzolo~\shortcite{mazzolo2014cauchy} has noted that, by reciprocity, the\remove{spectral decomposition of} \new{path-length spectrum for} flux leaving a boundary is identical to the distribution \remove{of free paths} for that flux finding its first collision in the medium in the adjoint direction.  This can be found by normalizing $p_u(s)$ from the boundary to the back of the region, and we conjecture that this will apply generally in GRT, \new{provided weak reciprocity holds for the subpath~\cite{deon2018reciprocal}}.  \new{This} will be required to embrace $\psi(\pos,\dir,s)$ directly,\remove{where required} \new{when needed}.}  Generalizing Placzek's Lemma~\cite{case53} and adding/doubling~\cite{vandehulst80,peltoniemi1993radiative} will also require the notion of virtual boundaries and $\psi(\pos,\dir,s)$.

  \removetwo{The study of Gamma-2 flights in three dimensions has yielded some of the first exact, analytic solutions for non-trivial scattering and absorption problems in a 3D domain.  These solutions add important contributions to the body of analytical benchmarks for Monte Carlo codes, and require no numerical integration or convergence acceleration to utilize.  It remains to further explore the Gamma-2 solutions in bounded and time-dependent scenarios.  The role of the Laplace transform in deriving time-dependent solutions will require generalization in GRT.  While the scalar collision-rate density is exactly a diffusion solution, the angular and time-dependent collision-rates are not (the density of first collisions about a point source is singular in angle and the time-dependent Green's function for collision-rate will not be a Gaussian distribution).  Such exploration should yield additional benchmarks and shed further light on the role of diffusion in GRT and will be explored in a following paper.}

  Close inspection of the relationship between flux and collision rate has highlighted some subtle relationships that are often taken for granted or, occasionally, misunderstood.  In light of that, there may be benefit in revisiting some classical settings with a new lens.  For example, the lengthy treatment of transmittance estimation in the short form of photon beams~\cite{jarosz2011progressive} can be collapsed to noting simply that the track-length estimator is an unbiased estimator for the \emph{flux} in heterogeneous classical media under delta tracking and that, needing the \emph{collision-rate}, not flux, to send photons towards the camera, requires the track-length-collision-rate estimator.  Provided that the density estimation kernel is small and the cross section is approximately constant across it, the cross-section integral can be approximated.  More generally, the role of Monte Carlo methods in computer graphics might benefit from being recast as fundamentally placing less importance on flux in scattering volumes and more importance on collision rate, moving away from a notion of transmittance estimation to one of collision-rate estimation.  Transmittance estimation seems only relevant in graphics at a camera sensor or on the surface of a light during next-event estimation, neither of which involve scoring along a track-length.

  There would appear to be great promise for solving plane-parallel problems with isotropic scattering in GRT by generalizing the Schwarzchild-Milne kernel for the scalar collision density to
  \begin{equation}
    K_C(x) = \frac{c}{2} \int_0^1 p_c(|x| / \mu) \frac{1}{\mu} d\mu,
  \end{equation}
  and solving a \new{Fredholm} integral equation of the second kind with a kernel that is no longer $E_1(|x|)$.  
    \new{The approach is then nearly identical to that of problems with complete frequency redistribution, and the Wiener-Hopf equations have been solved in the general case~\cite{mullikin1968some,ivanov1994resolvent}.}  
    In part 4 of this work we treat bounded plane-parallel problems in detail with a particular focus on the Gamma-2 free-path distribution.  Part \new{5} will explore approximate deterministic solutions of searchlight problems utilizing the method of images.

    \subsection{Limitations}
      \removetwo{\new{
      If we go back to the relationship between RGRT and wave theory and lift the initial assumption that scalar geometrical optics is sufficiently accurate in any specific realization, we raise the important open question: to what extent is RGRT related to honest wave propagation theories for correlated random media, and what, if any role, does Equation~\ref{eq:pcXu} play?  Under restrictive enough assumptions, classical exponential RT has been found to be a valid counterpart to wave theory~\cite{borovoi02,mishchenko2006radiative}.  However, a comprehensive wave theory must include the regime where geometrical optics is highly accurate, leading us to conclude that exponential RT will not be an appropriate counterpart to a more general theory, and to suspect that something like RGRT will.  Clarifying the relationship of the current work to a comprehensive wave theory remains an important and challenging open problem and will not be tackled presently.  In the mean time, the practitioner applying RGRT to wave propagation problems should be comfortable with the limitations of the ``motley''~\cite{mishchenko08} concepts of a Neumann series of collisions of a point photon traveling along piecewise-straight paths together with the limitations of the scalar approximation, particularly for coherent backscattering (even with unpolarized illuminants)~\cite{mishchenko2008accuracy}.}}

    The present assumption of path-length independence on single-scattering albedo limits the scope of application to random media where absorption and scattering occur from the same correlated structure.  This would not be an appropriate assumption of, say, metallic sphere packings in a paint layer, where absorption occurs in the substrate that holds the spheres, because the single-scattering albedo in this scenario would depend on $s$ (the equivalence theorem~\cite{vandehulst80} does not hold in GRT).  This calls into question the application of this form of GRT to tissue optics~\cite{wrenninge17} where scattering is primarily caused by random changes of index of refraction in the cellular tissue structures, while absorption is primarily caused by independent distributions of melanin and hemoglobin chromophores~\cite{wang2000modelling,tuchin07}.  We remark that it is straightforward to extend much of the current presentation to include correlated scattering that is imbedded in a classical, purely-absorbing substrate with absorption coefficient $\Sigma_a$.  An additional, reciprocal attenuation $\e^{-\Sigma_a s}$ is added to the Monte Carlo estimators to treat the substrate absorption in the expected-value manner.  Extension of the Fourier derivation of Green's function is less straightforward, however, and we leave this for future work.  \new{Extending the methods of the present paper to} a fully general GRT framework of path-length dependent absorption~\cite{rukolaine2016generalized} phase function~\cite{jarabo18} \new{and support for anomalous diffusion~\cite{liemert2017radiative}} would be a much larger task and the conditions for Helmholtz reciprocity in such settings are not immediately obvious.

    \removetwo{Further exploration of the relationships between explicit random structures and their \new{ensemble-average 2-point} free-path statistics would be beneficial from many perspectives.  \new{In the next paper in this series we consider stochastic binary mixtures in infinite spherical and plane geometries with nonstochastic reaction rates and nonstochastic albedo.  Adopting the Levermore-Pomraning attenuation law as $X_u(s)$ in Table~\ref{tab:notation} produces the condensed-history acceleration over phase transitions proposed and studied by Audic and Frisch, without the requirement for the histogram construction of $p_c(s)$.  We derive some exact and approximate deterministic solutions using a single convolution integral equation over collision rate density as an alternative to the coupled equations required when treating the fluxes directly.  Multi-dimensional benchmark solutions} provide means to evaluate the limitations of the 2-point statistics and to compare to the chord-length sampling algorithms for binary mixtures.  Benchmarks focusing on collision rate and stochastic reaction rate seem largely missing, however.}

      There is also important open work to relate a given model in GRT to a scale at which it can be safely assumed that no explicit structure would be resolved in imaging that media at that distance, making the bulk statistical model an appropriate choice.

      A comprehensive notion of heterogeneity in GRT (non-uniform mean density) \removetwo{in non-continuum models} is also an interesting open question, given that optical path-length stretching~\cite{bitterli2018radiative} would not be able to avoid also stretching density-invariant aspects of correlation.  Consider, for example, monodisperse hard sphere packings in a medium where the average number density of spheres changes continuously with position.  The restriction of non-intersection is not a function of the local number density and does not scale in the sense of the traditional optical depth.

      Invariant properties of general random walks~\cite{frisch1995universality,mazzolo2014cauchy} provide additional benchmarking tools for both Monte Carlo and deterministic methods.  Under the mental model for GRT described in Section~\ref{sec:mentalmodel} it seems intuitive that the beginning of an uncorrelated free path involves all random realizations in equilibrium and that, therefore, the macroscopic cross-section beginning that walk should be related to the number densities $\rho_i$ and constituent cross-sections $\sigma_i$ in the classical \new{atomic-mix} sense, independent of correlation,
      \begin{equation}
        \Sigma_{tu}(0) = \sum_i \rho_i(\pos) \sigma_i.
      \end{equation}
      We have observed that this property holds in the blue noise GRT model previously proposed~\cite{deon2018reciprocal}, in Davis's power-law model~\cite{davis2011radiation}, \new{and for Markovian binary mixtures (presented in the next paper)}, but, without further evidence, we leave this as a conjecture for now.

\section{Conclusion}

  We have presented new Monte Carlo and deterministic methods for nonclassical linear transport and related stochastic processes.  The collision estimator for collision rate and the track-length estimator for flux both have been shown to remain equivalent to the classical case, with each estimator requiring path-length-dependent scores when estimating the opposite quantity (summarized in Figure~\ref{fig-gen-sigmat}).  Limitations of the collision estimator for flux estimation in negatively-correlated media have been identified and a new family of estimators based on fictitious scattering has been introduced to mitigate these limitations.

  The Monte Carlo estimators have been verified using new Green's function derivations for point and plane sources in infinite media, extending Grosjean's random flight solutions to include the transmittance kernel for flux and a unique PDF for uncorrelated-origin free paths in a unified manner.  Algebraic equations for the related moment-preserving diffusion approximations have also been derived, requiring moments up to order $4$ in some cases.  We have presented the first generalized moment-preserving diffusion asymptotics for scalar flux and have shown that previous asymptotics correspond to classically-scaled asymptotics for the collision rate.

\section{Acknowledgements}

  \new{I am indebted to M.M.R. Williams for introducing me to the work of Larsen and Vasques and for his encouragement of the present work.  Discussions of transport and memory with Barry Ganapol, Wenzel Jakob, Steve Marschner and Anthony Davis were also very helpful.  I also need to thank Todd Urbatsch for pointing me to the work of Fleck and Canfield~\shortcite{fleck84}, Eric Heitz for mentioning the work of Savo et al.~\shortcite{savo2017observation}, Andrea Zoia for his thorough review of an early revision and the anonymous referees for their help improving the presentation.}

\bibliographystyle{acmsiggraph}
\bibliography{nonexppartii}

\appendix

\section{Gamma-2 Flux Derivation in 3D}\label{appendix:gamma2flux}

  The plane-geometry kernel $K_\phi(x)$ producing collided flux from the collision rate density in 3D transport with isotropic scattering is
  \begin{equation}
  	K_\phi(x) = \frac{c}{2} \int_0^1 X_c(|x| / \mu) \frac{1}{\mu} d\mu,
  \end{equation}
  which for Gamma-2 transport is
  \begin{equation}
  	K_\phi(x) = \frac{c}{2} \left( \e^{-|x|} + E_1(|x|) \right).
  \end{equation}
  Mathematica is able to compute the collided flux from a collision rate density given by a diffusion mode $C(x) = \e^{-|x| / \nu}$,
  \begin{multline}\label{eq:kphiplane}
  	\int_{-\infty}^\infty \e^{-|t| / \nu} K_\phi(t-x) dt = c \nu  E_1(\left| x\right| ) \\
  	 + \frac{c \nu  e^{-\frac{\left| x\right| }{\nu }} \left(-e^{\left(\frac{1}{\nu }-1\right)
   \left| x\right| }+\left(\nu ^2-1\right) \coth ^{-1}(\nu )+\nu \right)}{\nu ^2-1} \\
   +\frac{1}{2} c \nu  e^{-\frac{\left| x\right| }{\nu }}
   \left[-E_1\left(\left(1-\frac{1}{\nu }\right) \left| x\right| \right)-e^{\frac{2
   \left| x\right| }{\nu }} E_1\left(\frac{(\nu +1) \left| x\right| }{\nu
   }\right)\right].
  \end{multline}
Performing the convolution in Eq.~\ref{eq:kphiplane} in the frequency domain using
  \begin{equation}
  	\int_{-\infty}^\infty \e^{-|t| / \nu} \e^{i z x} dx = \frac{2 \nu}{1 + \nu^2 z^2}
  \end{equation}
  and
  \begin{equation}
  	\int_{-\infty}^\infty K_\phi(x) \e^{i z x} dx = c \left(\frac{1}{z^2+1}+\frac{\tan ^{-1}(z)}{z}\right)
  \end{equation}
  we find that the convolution is given by the following Fourier inversion
  \begin{multline}
  	\int_{-\infty}^\infty \e^{-|t| / \nu} K_\phi(t-x) dt \\= \frac{1}{2\pi} \int_{-\infty}^\infty \frac{2 \nu}{1 + \nu^2 z^2} c \left(\frac{1}{z^2+1}+\frac{\tan ^{-1}(z)}{z}\right) \e^{-i z x} dx.
  \end{multline}
  Given that the convolution of the two diffusion terms is simple
  \begin{equation}
  	\frac{1}{2\pi} \int_{-\infty}^\infty \frac{2 \nu}{1 + \nu^2 z^2} c \left(\frac{1}{z^2+1}\right) \e^{-i z x} dx  = \frac{c \nu  \left(e^{-\left| x\right| }-\nu  e^{-\frac{\left| x\right| }{\nu }}\right)}{1-\nu ^2}
  \end{equation}

  we can solve to find that the Fourier inversion of the arctan term is therefore
  \begin{multline}
  	Z_{pl}(x,\nu) = \frac{1}{2\pi} \int_{-\infty}^\infty \frac{1}{1 + \nu^2 z^2}  \left(\frac{\tan ^{-1}(z)}{z}\right) \e^{-i z x} dx \\ = \frac{1}{2} \coth ^{-1}(\nu ) e^{-\frac{\left| x\right| }{\nu }} + \frac{E_1(\left| x\right| )}{2} -\frac{1}{4} e^{-\frac{\left| x\right| }{\nu }} E_1\left(\left(1-\frac{1}{\nu }\right) \left| x\right|
   \right) \\ -\frac{1}{4} e^{\frac{\left| x\right| }{\nu }} E_1\left(\frac{(\nu +1) \left| x\right| }{\nu }\right).
  \end{multline}
  Using the plane-to-point transform~\cite{case53}
  \begin{equation}
	    \phipt(r) = -\frac{1}{2 \pi r} \frac{d \phi_{\text{pl}}(x)}{dx} \vert_{x=r}
	  \end{equation}
  we also have the spherical-geometry equivalent,
  \begin{multline}
  	Z_{pt}(r,\nu) = \Finvthree \left\{ \frac{\tan^{-1} z}{z} \frac{1}{1 + (z \nu)^2} \right\} = \\ \frac{e^{-\frac{r}{\nu }} \left(-E_1\left(r \left(1-\frac{1}{\nu
   }\right)\right)+e^{\frac{2 r}{\nu }} E_1\left(r \left(1+\frac{1}{\nu }\right)\right)+2
   \coth ^{-1}(\nu )\right)}{8 \pi  \nu  r}.
  \end{multline}
  With these results we are able to express the scalar flux about the correlated-emission isotropic plane source (Eq.~\ref{eq:phicplane})
  \begin{equation}
  	\phi_c(x) = \frac{E_1(\left| x\right| )}{2} + \frac{e^{-\sqrt{1-c} \left| x\right| }}{2 \sqrt{1-c}}
          \, + \frac{c}{1-c} \, Z_{pl}(x,\frac{1}{\sqrt{1-c}})
  \end{equation}
  and the scalar flux about the correlated-emission isotropic point source, Eq.~\ref{eq:gammaphic}
  \begin{equation}
  	\phi_c(r) = \frac{\e^{-r}}{4 \pi r^2} + \frac{\e^{-r \, \sqrt{1-c}}}{4 \pi r} + \frac{c}{1-c} Z_{pt}(r,\frac{1}{\sqrt{1-c}})
  \end{equation}

\section{Nonexponential Monte Carlo Estimation: Additional Details}\label{appendix:MC}

\subsection{Collision / Track-Length Duality and Joint Estimation}

      In Section~\ref{sec:MC}, we found two complementary pairs of estimators, two of the collision type and two of the track-length type.  In some sense, the collision estimator most naturally estimates collision rate, and the track-length estimator most naturally estimates flux.  This echoes the duality of flux creating collisions and collisions creating flux in linear transport theory.  Each estimator can compute its primal quantity locally in the smaller phase space with no memory (track-length for flux and collision rate for collision), while requiring the memory variable $s$ to compute the complementary quantity.  The advantages of both can be realized over the same random walk by having each contribute separately to distinct flux-proportional and collision-proportional tallies.

      It is natural to ponder whether or not there exist variants of each estimator that estimate the other quantity in a memory-free way, and indeed there are (with a few restrictions).  This can be achieved by sampling free-path lengths from non-analog PDFs, and essentially moves the memory-dependent calculation to a \emph{particle weight} $W$, which still ends up in the score.  This would seem to offer little advantage in the case of the collision estimator, but is more interesting in the case of track-length estimation.

      \new{For completeness, and to begin with the simpler case, let us first consider how this would work for the collision estimator}.  Consider correlated free-path lengths that are sampled from $p_u(s)$ instead of $p_c(s)$.  Collisions will now occur in proportion to the flux, due to the definition of $p_u(s)$.  A collision estimator scoring $W \s$ is then a memory-free unbiased collision estimator of the flux.  The weight factor begins the walk at $W = 1$ and updates prior to each collision using
      \begin{equation}
        W \leftarrow \left( W \frac{p_c(s)}{p_u(s)} = W \frac{\Sigma_{tc}(\pos,\dir,E,s)}{\s} \right).
      \end{equation}
      It is clear that we have simply moved the memory-dependent \new{term} $\Sigma_{tc}(s)$ into the weight factor \new{$W$}.

      For the case of track-length estimation of collision-rate, Heitz and Belcour~\shortcite{heitz18note} noted that sampling track-lengths proportional to the derivative of the free-path distribution (provided it is monotonically decreasing) and applying a normalization weight (a constant) to each track-length estimate is an unbiased estimator for the free-path distribution and, therefore, the collision rate density.  Again, the memory calculation is moved into a weight factor update at each such non-analog free-path sampling, but now the advantage is that the score along the track-length is a constant times the length and the integral of the cross-section is avoided.

      Employing either of these path-length biasings successively in creating a full random walk is inadvisable, due to the weight fluctuations that could occur and chain together.  However, a modified walk of the kind described by Sweezy~\shortcite{sweezy2018monte} could be adapted here as follows.  An analog walk is sampled as above, but its track lengths and collision sites are not used to score.  Instead, at each source vertex and non-terminating collision vertex of the analog walk, $N$ scoring tracks, each with weight \newtwo{factors} $1 / N$ are sampled using the phase function and non-analog free-path distributions just described (for the corresponding origin type).  By spawning the scoring tracks from the vertices of the analog walk, we both solve the same integral equation as would be solved by scoring from the walk itself, and contributions of all scattering orders are included.  Accumulated weight-factor fluctuations are, however, avoided by building the primary walk with analog free-path sampling.  Applying this method for track-length estimation of collision rate involves temporary weight factors for each scoring track, which depend on $s$ and scale the track-length scores.  These weights are then discarded after scoring and the walk continues in the analog fashion to the next vertex.

      The same idea can extend the collision-rate-density-estimation methods of the track-length type called ``photon beams'' by Jarosz et al.~\cite{jarosz2011progressive,novak18} for GRT.  These algorithms sample and store a number of particle histories (in the forward direction) into a data structure that is later used to share these calculations over millions of next-event estimations through a virtual camera aperture and onto the pixels comprising an image of light scattering within participating media, such as fog or tissue.  The next-event calculations are of the back-projection angular flux type~\cite{dunn1985back} and are combined with biased kernel density estimation.  The extension just described would yield memory-free track-length estimates across each density-estimation kernel, avoiding the cross-section integrals otherwise required to estimate the density of collisions that send light into the direction of the camera's aperture.

\subsection{Non-Analog Games}

         It is often advantageous to deviate from the analog walk by altering the probabilities of absorption, scattering angle, and free-path length sampling to reduce variance and improve performance~\cite{lux91,hoogenboom08b}.  \remove{Those}\new{The} most common and basic of such schemes utilize the classical particle weight concept, with weight corrections calculated in the usual fashion.

         Implicit capture \new{in GRT} remains unchanged, multiplying the particle weight $W$ by $c(\pos)$ at each collision and applying additional weight changes if Russian roulette is used to terminate the walk.  Collision and track-length scores are then scaled, per collision and track segment, by the current particle weight $W_i$.  Likewise, angle biasing can use non-analog phase functions $P'$ by adopting a weight factor
         \begin{equation}
          W \leftarrow W \times \frac{P(\dir_i,\dir_{i+1})} {P'(\dir_i,\dir_{i+1})}
         \end{equation}
         prior to the next score.  Path-length biasing from a non-analog PDF $p'(s)$ requires a similar correction for the collision estimators prior to scoring
         \begin{equation}
          W \leftarrow W \frac{p_\_(s)}{p'(s)}.
         \end{equation}
         For non analog free-path sampling $p'(s)$ (other than the forms previously discussed), the track-length estimators both become memory dependent and require changes, similar in derivation to the case presented above, which we skip for brevity.

      \subsection{Estimation at a Point}

        Next-event estimation (NEE) at a point in GRT takes on four forms, generalizing the classically equivalent estimators.  In all cases the geometry factor of $1 / r^2$ leads to an estimator of unbounded variance in 3D space~\cite{kalos63b} and should be avoided using equiangular sampling or similar approaches that remove the singularity~\cite{kalli77,rief84,kulla12,georgiev2013joint}.  The collision-rate expected value estimate at $\pos$ from a correlated-origin walk vertex $\pos_c$ is
        \begin{equation}
          \frac{p_c(||\pos - \pos_c||) P(\dir_i,\frac{\pos - \pos_c}{||\pos - \pos_c||})}{||\pos - \pos_c||^2}.
        \end{equation}
        For uncorrelated-origins, $p_c$ is replaced with $p_u$.  For flux expected value estimates, the free-path distributions $p_c(s)$, and $p_u(s)$ change to transmittances $X_c(s)$ and $X_u(s)$ respectively for correlated and uncorrelated origins.  Integration of these estimators along track-lengths yields the respective generalizations.

      \subsection{Dwivedi Guiding}

        Analog walks are particularly ineffective at estimating extremely low values, such as shielding calculations through very thick slabs.  A highly effective method for guiding the walk in this case toward a zero-variance estimator for penetration through the slab is to assume an approximate importance function of escape at each optical depth $z$ with translational invariance of the form~\cite{dwivedi82}
        \begin{equation}
          I_e(z) = \e^{-z/\nu_0}
        \end{equation}
        where $\nu_0 > 1$ is the largest discrete eigenvalue for the transport kernel~\cite{case67}.  We found this approach to generalize for GRT, provided, at least, $p_c(s)$ has bounded moments of all orders, by noting that angle selection, assuming an exponential escape probability for exiting a plane-parallel system after entering a collision at depth $z$, is the normalization of
        \begin{equation}
          \phi_{\nu_0}(\mu) = \frac{1}{2} c \int_0^{\infty } p_c(t) \exp \left(\frac{\mu}{\nu_0} t\right) \, dt,
        \end{equation}
        \new{where $\mu$ is the direction cosine to the depth axis.}
        For the Gamma-2 free path distribution $p_c(s) = \e^{-s} s$ we find
        \begin{equation}
          \phi_{\nu_0}(\mu) = \frac{c}{2} \left( \frac{\nu_0}{\nu_0 - \mu} \right)^2,
        \end{equation}
        which is normalized
        \begin{equation}
          \int_{-1}^1 \phi_{\nu_0}(\mu) d\mu = 1
        \end{equation}
        provided $\nu_0 = 1 / \sqrt{1-c}$.  We found it practical to sample free-paths and angles using this generalized Case eigenmode and found a five order of magnitude reduction in variance for the transmission of normally-incident beam illumination through a 20 mean-free-path thick slab with single-scattering albedo $c = 0.7$ compared to the analog walk.  We \new{postpone} the details, in part, because in Part \new{4} of this work we present the perfectly zero-variance walk for the half space related to this approach, which will be a better forum for highlighting how the Dwivedi method changes in GRT.

        For the power-law free-path distributions discussed above, the normalization of this angular importance distribution is impossible for $\mu / \nu_0 > 0$, so the method would only apply to angle selection in one hemisphere.  In the directions where the exponential importance increases, the integral of $p_c(s) I_e(z(s))$ is unbounded.  Clamping $I_e(z)$ outside of the slab would be required, which complicates the sampling procedure, but may be practical in some cases. 

  \subsection{Sampling Free-Path Distribution Tails}\label{sec:tails}
    To sample the tail of some PDF $p(s)$ starting at $p_0$, where the sampling procedure for all of $p(s)$ used the single-random-variable CDF-inversion trick, we can simply remap the range of the random number using the CDF to exclude the range $[0,p_0]$, which is necessarily possible because we needed to know that CDF to invert it for the standard sampling procedure.  This yields a range of uniform random number $\xi \in [a,1]$ (or possibly $\xi \in [0,a]$), which, when passed to the sampling procedure for $p(s)$, yields $s \in [p_0,\infty]$, from which we subtract $p_0$.

    \new{For free-path distributions, such as Mittag-Leffler, where the CDF inversion method is not possible~\cite{fulger2008monte}, alternative methods will be required.}

\end{document}